\documentclass[useAMS,usenatbib,usegraphicx]{aa}
\usepackage{txfonts,graphicx}
\usepackage[normalem]{ulem}
\usepackage{natbib}
\usepackage{color}

\newcommand{\HI}{{\rm H\,\scriptstyle I}}
\newcommand{\HII}{{\rm H\,\scriptstyle II}}

\newcommand{\dd}{\mathrm{d}}

\newcommand{\cm}{\,{\rm cm}}

\newcommand{\kpc}{\,{\rm kpc}}
\newcommand{\Mpc}{\,{\rm Mpc}}

\newcommand{\yr}{\,{\rm yr}}

\newcommand{\radm}{\,{\rm rad\,m^{-2}}}

\definecolor{webgreen}{rgb}{0,.5,0}
\definecolor{webbrown}{rgb}{.6,0,0}
\definecolor{purple}{rgb}{0.5,0,.5}

\begin{document}

\title{Magnetic and gaseous spiral arms in M83}

\author{P. Frick\inst{1}, R. Stepanov\inst{1}, R. Beck\inst{2}
\thanks{Corresponding author: \email{rbeck@mpifr-bonn.mpg.de}} ,
D. Sokoloff\inst{3}, A.~Shukurov\inst{4}, M.~Ehle\inst{5}, and A.~Lundgren\inst{6}}
\titlerunning{Magnetic and gaseous arms in M83}
\authorrunning{P. Frick et al.}
\institute{Institute of Continuous Media Mechanics, Korolyov str. 1, 614061 Perm, Russia \and MPI
f\"ur Radioastronomie, Auf dem H\"ugel 69, 53121 Bonn, Germany \and Department of Physics, Moscow
State University, Moscow, 117588, Russia \and School of Mathematics and Statistics, Newcastle
University, Newcastle upon Tyne NE1~7RU, U.K. \and {\em XMM-Newton} Science Operations Centre,
ESAC, ESA, PO Box 78, 28691 Villanueva de la Ca\~nada, Madrid, Spain \and Joint ALMA Observatory,
Alonso de Cordova 3107, Santiago, Chile}



\abstract{The magnetic field configurations in several nearby
spiral galaxies contain magnetic arms that are sometimes located between the material arms.
The nearby barred galaxy M83 provides an outstanding example
of a spiral pattern seen in tracers of gas and magnetic field.}
    {We analyse the spatial distribution of magnetic fields in M83 and their relation to the material
spiral arms.
}
    {Isotropic and anisotropic wavelet transforms are used to
decompose the images of M83 in various tracers to quantify structures in a range of
scales from 0.2 to 10\kpc. We used radio polarization observations at $\lambda6.2\cm$ and
$\lambda 13$\,cm obtained with the VLA, Effelsberg and ATCA telescopes
and APEX sub-mm observations at $870\,\mu$m, which are first published here,
together with maps of the emission of warm dust, ionized gas, molecular gas, and atomic gas.}
        {The spatial power spectra are similar for the tracers of
dust, gas, and total magnetic field, while the spectra of the
ordered magnetic field are significantly different. As a consequence, the wavelet cross-correlation between
all material tracers and total magnetic field is high, while the structures of the ordered
magnetic field are poorly correlated with those of other tracers. The magnetic field configuration
in M83 contains pronounced magnetic arms. Some of them are displaced from the corresponding
material arms, while others overlap with the material arms. The pitch angles of the magnetic
and material spiral structures are generally similar.
The magnetic field vectors at $\lambda6.2\cm$ are aligned with the outer material
arms, while significant deviations occur in the inner arms and, in particular, in the bar region,
possibly due to non-axisymmetric gas flows.
Outside the bar region, the typical pitch angles of the material and magnetic spiral arms are
very close to each other at about $10^\circ$.
The typical pitch angle of the magnetic field vectors is about $20^\circ$ larger
than that of the material spiral arms.}
        {One of the main magnetic arms in M83 is displaced from the gaseous arms similarly to the
galaxy NGC\,6946, while the other main arm overlaps a gaseous arm,
similar to what is observed in M51.
We propose that a regular spiral magnetic field generated by a mean-field dynamo is compressed
in material arms and partly aligned with them.
The interaction of galactic dynamo action with a transient spiral pattern is a promising
mechanism for producing such complicated spiral patterns as in M83.}

\keywords{magnetic fields -- MHD -- galaxies: ISM -- galaxies: individual: M83 -- galaxies:
magnetic fields -- galaxies: spiral}

\maketitle

\section{Introduction}                             \label{Int}

\begin{figure}
\vspace{25mm}
\centering
\includegraphics[width=0.9\columnwidth]{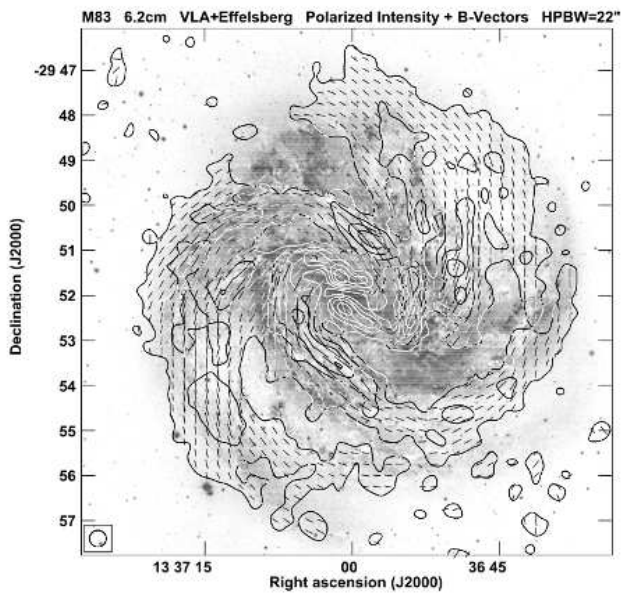}
\caption{Polarized radio continuum intensity (contours) and $B$ vectors of M83 at $\lambda6.17\cm$,
obtained by combining data from the VLA and Effelsberg telescopes, overlayed on an optical image
from the Anglo-Australian Observatory by Dave Malin. The angular resolution is $22\arcsec$.
Faraday rotation of the $B$ vectors has not been corrected, but is small because of the short
wavelength and the low inclination of the galaxy. }  \label{fig:pi6}
\end{figure}

\begin{figure}
\vspace{25mm}
\centering
\includegraphics[width=0.9\columnwidth]{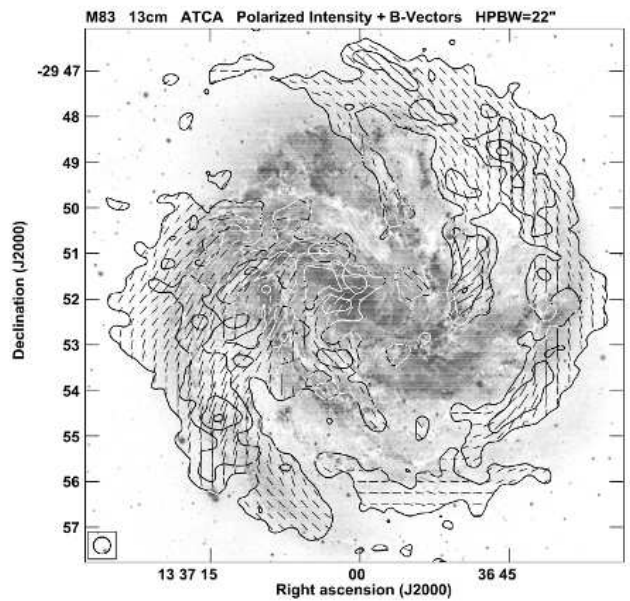}
\caption{Polarized radio continuum intensity (contours) and $B$ vectors of M83 at $\lambda12.8\cm$,
observed with the ATCA telescope and overlayed onto an optical image from the Anglo-Australian
Observatory by Dave Malin. The angular resolution is $22\arcsec$. Comparison with
Fig.~\ref{fig:pi6} shows that the Faraday rotation of the $B$ vectors is significant at this
wavelength. }  \label{fig:pi13}
\end{figure}

M83 (NGC~5236) is the nearest barred galaxy and is well studied at all wavelengths, including
polarized and unpolarized radio continuum emission that traces the interstellar magnetic field.
Early polarization observations of M83 were performed by \citet{Sukumar87} at the wavelengths
$\lambda92\cm$, $\lambda20\cm,$ and $\lambda6.3\cm$ at a resolution of $33\arcsec\times55\arcsec$,
which corresponds to about $1.4\times2.4\kpc$ at the assumed distance to the galaxy of 8.9\,Mpc
\citep{Sandage74} \footnote{The distance to M83 is not well known; distance estimates listed in
the NASA/IPAC Extragalactic Database (NED) range between 4.5\,Mpc and 14.6\,Mpc .}, so that
$1\arcmin\approx2.6\kpc$. Significant polarized emission has been detected, indicating that the
galaxy hosts an ordered magnetic field approximately aligned with the spiral arms.
\citet{Allen89,Sukumar89} observed the galaxy with the VLA at $\lambda20\cm$ at a resolution of
$30\arcsec\times70\arcsec$ and found that the regions where polarized emission is strongest are located
outside the prominent optical spiral arms. However, the resolution of these observations was too
low and Faraday depolarization at this wavelength was too strong to reach firm conclusions about
the spatial distribution of the ordered magnetic field and its relation to the material spiral
pattern. \citet{Neininger91} observed M83 at $\lambda2.8\cm$ and found ordered fields along the
central bar and between the inner spiral arms. The Faraday rotation measures between these
wavelengths indicated a bisymmetric field structure possibly driven by the bar \citep{Neininger93}.
More recent observations at $\lambda12.8\cm$ with the ATCA data
show the radio continuum emission from the bar and spiral arms in detail (Figs.~\ref{fig:pi6} and
\ref{fig:pi13}).

Our aim is to determine the parameters of the spiral arms in M83, as seen in different tracers,
and to study their morphological relations. We are particularly interested in the relative
positions of gaseous and magnetic arms, especially those of the large-scale magnetic field.
\citet{Hoernes96} discovered that the ordered magnetic field in the galaxy NGC\,6946, traced by
polarized radio continuum emission, concentrates in well-defined magnetic arms, which are
interlaced with the gaseous arms. The comparative morphology of the spiral patterns visible in
different tracers was discussed by \citet{frick00} who confirmed the conclusions of
\citet{Hoernes96} and determined such parameters as the pitch angle, width, and arm--interarm
contrast for the magnetic arms and those seen in optical red light. A remarkable feature of the
spiral pattern in NGC\,6946 is that the phase shift between the magnetic and stellar/gaseous arms is
more or less constant with radius. In other words, the two spiral arm systems have similar pitch
angles and do not intersect in the part of the galaxy explored.

The phenomenon of magnetic arms may be common among spiral galaxies. It was first observed in the
spiral galaxy IC\,342 by \citet{Krause89} \citep[see also][]{Krause93}. However, the relation
between magnetic and gaseous arms in IC\,342 \citep{beck15} and other galaxies is not as simple and
clear-cut as in NGC\,6946.
Another example is provided by M51 where the ordered magnetic field is maximum at the spiral arms
traced by dust lanes in some regions and displaced from them at other locations \citep{fletcher11}.
When applying a wavelet analysis with anisotropic wavelets, the pitch angle of the spiral arms in M51
was found to vary with the galacto-centric radius \citep{patrikeyev06}. Systematic shifts between
the spiral arm ridges in gas, dust, and magnetic field indicate a time sequence that is consistent with the
gaseous spiral arms causing compression that results in a stronger magnetic field.


To assess the origin, role, and significance of magnetic arms and their relation to
material arms, one needs a sample of galaxies with different spiral patterns. This paper adds one
remarkable galaxy to this sample. We performed a systematic analysis of the spiral patterns in M83
using ancillary data and new observations and isotropic and anisotropic wavelet analysis similar
to that of \citet{frick01} and \citet{patrikeyev06}.
Our quantitative methods are free of restrictive ad hoc assumptions,
such as logarithmic spirals with a fixed number of arms and constant pitch angles.
To make our analysis more comprehensive, we used polarized radio continuum observations
at $\lambda6\cm$ (Fig.~\ref{fig:pi6}) and $\lambda13\cm$ (Fig.~\ref{fig:pi13}), which are first
published here.

The main message from the analysis performed here is that the configuration of magnetic and
material arms in M83 is partly similar to M51. In both galaxies, the magnetic arms overlap
the material pattern in large areas but not everywhere. Of course, the numbers of arms, their pitch
angle, and other quantitative parameters are specific for each galaxy.

\section{The data}                            \label{TD}

\subsection{ATCA radio continuum observations at $\lambda13\cm$}
\label{ATCA}

\begin{figure}
\vspace{4mm}
\centering
\includegraphics[width=0.95\columnwidth]{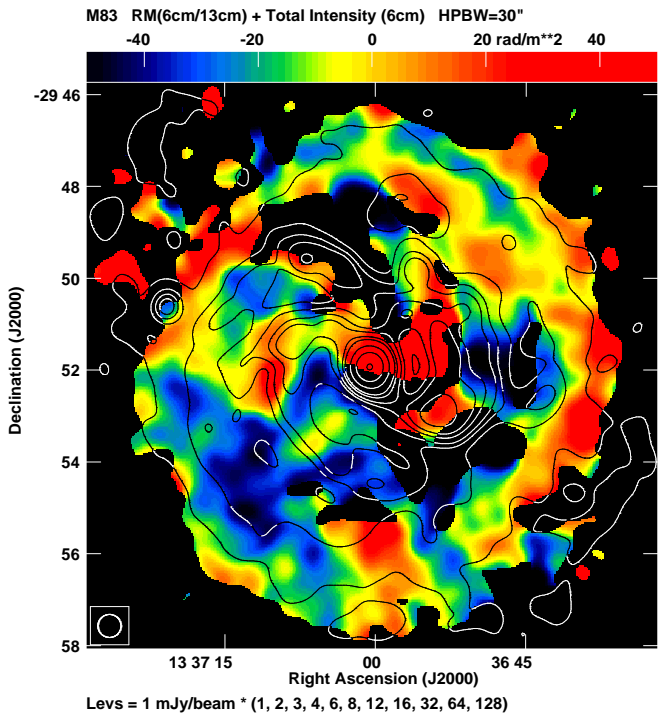}
\caption{Faraday rotation measures between $\lambda6.17\cm$ and $\lambda12.8\cm$ (colours) and
total radio continuum intensity at $\lambda6.17\cm$ (contours). The angular resolution is $30\arcsec$.}
\label{fig:rm}
\end{figure}

We performed radio continuum observations of M83 with the Australia Telescope Compact Array
(ATCA) \footnote{The Australia Telescope Compact Array is part of the Australia Telescope National
Facility, which is funded by the Commonwealth of Australia for operation as a National Facility
managed by CSIRO.}, consisting of six parabolic dishes, each of 22\,m in diameter, forming an
east-west radio interferometer.

We observed M83 with different baseline configurations (Table~\ref{arrays}), each observation
session lasting 12\,hours to achieve maximum $uv$ coverage. The array was used in its
five-telescope compact configuration ignoring the longest baselines (i.e., combinations with the
Antenna~6) that are most prone to phase instabilities. The names of the configurations used in the
table give approximately the largest baseline (750\,m) used for this project. The full width at
half maximum (FWHM) of the primary beam at $\lambda13\cm$ is $22\arcmin$, large enough to map the
radio emission of M83 with an angular size of about $15\arcmin$ at $\lambda 20\cm$
\citep{Neininger93}. Also, the sampling of the shortest baselines is high enough to map the
extended radio emission.

Each antenna receiver was used in the 128\,MHz bandwidth continuum polarization mode. This
bandwidth consists of 32 separately sampled channels. Linear polarization was measured using
the orthogonal $X$ and $Y$ feeds on each antenna. Two frequencies (IFs) and
the two polarization channels
at each frequency were recorded simultaneously with 15\,s time resolution.

For all observations the phase centre was set at $\rm RA = 13^\mathrm{h}37^\mathrm{m}00\fs3,\  Dec
= -29\degr 53\arcmin 04\arcsec$ (J2000), which included an offset of 1\arcmin\ south of the
galaxy's center, so as to place the observational phase center away from the bright nuclear
emission of M83. This was done to check for systematic phase errors in the synthesis array that
would be visible around the phase centre.

To calibrate the observations of M83, the phase calibrator QSO J1313$-$333 (unresolved at our
baselines) was observed every half hour for about 5\,min. The primary calibrator PKS~1934$-$638,
observed for a short time during each synthesis run, was used as an absolute flux density
calibrator to which all other measurements were scaled.

\begin{figure*}
\centering
\includegraphics[width=0.85\textwidth]{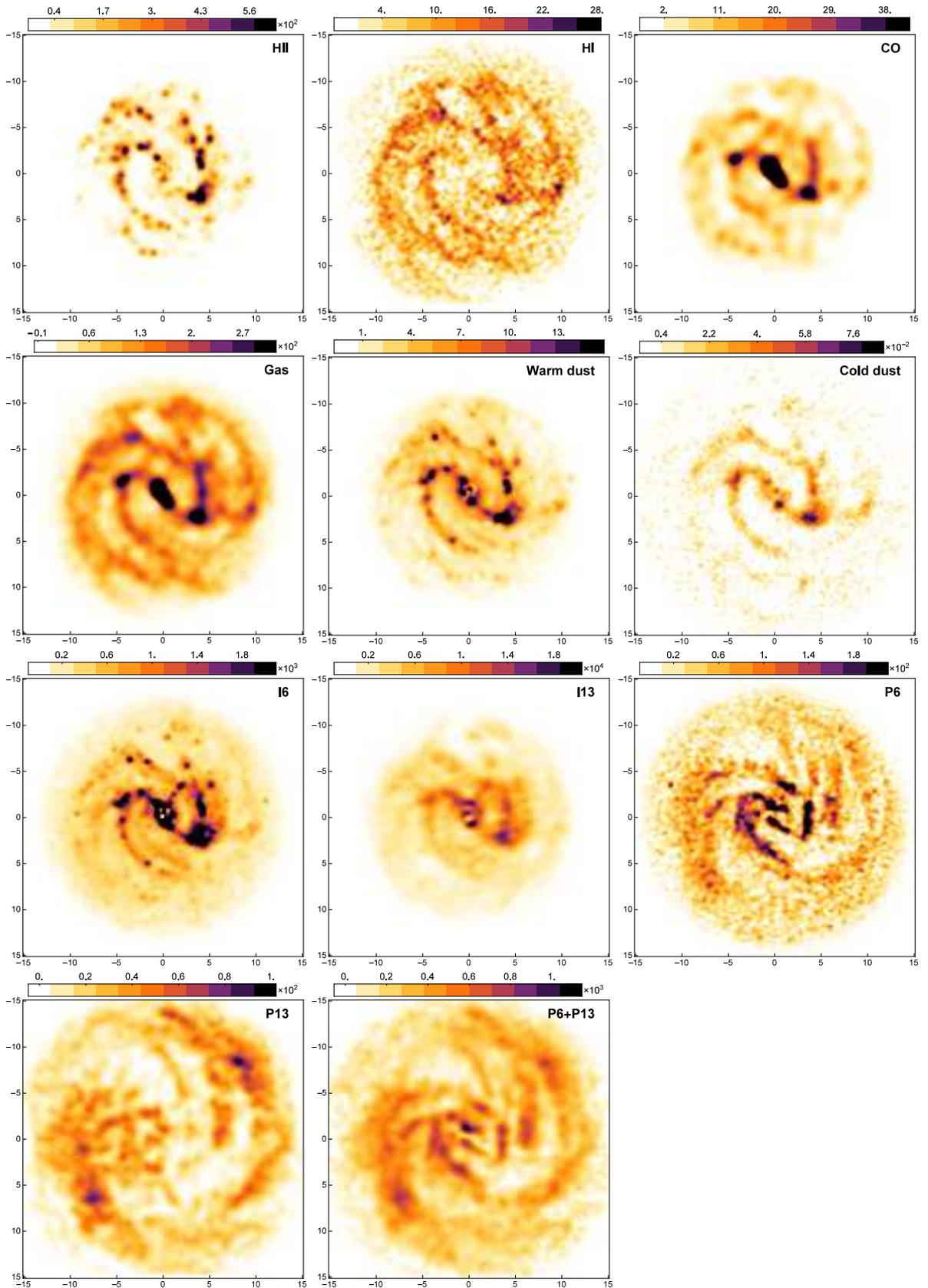}
\caption{\label{fig:maps}Maps of various tracers in M83: material tracers ($\HII$, $\HI$, CO, total
neutral gas, warm dust, and cold dust) and tracers of the total magnetic field (mainly its
small-scale part) (I6 and I13) and its ordered component (P6, P13, and P6+P13), as indicated in the
upper right corner of each frame. The coordinates are given in kpc, the intensity units are
arbitrary units ($\HII$), $10^{20}$\,cm$^{-2}$ ($\HI$), K~km/s (CO), $10^{20}$\,cm$^{-2}$ (gas),
MJy/sterad (warm dust), Jy/beam (cold dust), $\mu$Jy/beam (radio I6, I13, P6, P13, P6+P13). The
axis scales are in kpc.}
\end{figure*}

Data processing was done with the AIPS software package. The data of linearly polarized intensity were
calibrated within the MIRIAD (Multichannel Image Reconstruction Image Analysis and Display)
package. The calibrated visibility data of M83 were converted into maps of the Stokes parameters
$I$, $Q$, and $U$. The self-calibration has improved the total-power image of Fig.~\ref{fig:maps}
(Panel 8). The $Q$ and $U$ maps were combined into a map of polarized intensity $P=(Q^2+U^2)^{1/2}$
(Fig.~\ref{fig:pi13} and Panel 10 of Fig.~\ref{fig:maps}). The positive bias due to rms noise has
been corrected for the inner parts of the galaxy where the noise is smallest and almost constant.

The polarized emission at $\lambda13\cm$ is strongest in the outer regions of the galaxy. The
typical degrees of polarization are 2\% in the central region and 30\% in the outer galaxy. In
contrast to $\lambda13\cm$, the polarized emission at $\lambda6.2\cm$ (Fig.~\ref{fig:pi6}) is
strongest in the central region, confirming that Faraday depolarization is strongest in this region
\citep{Neininger93}. In both maps, the rms noise increases with increasing distance from the map's
centre.

Comparison of the position angles $\chi$ of the $B$ vectors between Figs.~\ref{fig:pi6} and
\ref{fig:pi13} reveals a generally clockwise rotation at $\lambda13\cm$. We computed classical
Faraday rotation measures, defined as $\mathrm{RM}=\Delta\chi\,/\,(\lambda_2^2-\lambda_1^2)$ (Fig.~\ref{fig:rm}).
Classical RMs are not reliable in the inner part of the galaxy ($< 3\arcmin$ radius) where
Faraday depolarization is strong at $\lambda13\cm$. In the outer disk, RMs vary between about
$-60\radm$ and $+60\radm$. No systematic variation RM with azimuthal angle, a signature
of a large-scale regular magnetic field in the disk, as in M51 \citep{fletcher11}
and in IC\,342 \citep{beck15}, and no traces of spiral arms are found,
which is not surprising in view of the small magnetic field component of the disk field
along the line of sight in a weakly inclined galaxy like M83. Furthermore, the strong
depolarization at $\lambda13\cm$ in the inner region leads to deviations of Faraday rotation angle
$\Delta\chi$ from the $\lambda^2$ law,
so that classical RMs are not reliable and wide-band spectro-polarimetric data and
application of RM Synthesis \citep{brentjens05} are required to measure the large-scale fields.

Still, the average rotation measure of $\simeq -11 \pm 18 \radm$ is useful as a measure of
the rotation in the foreground of our Milky Way. The corresponding rotation angle is $\simeq
-2\degr$ at $\lambda6.2\cm$ and $\simeq -10\degr$ at $\lambda13\cm$.
According to \citet{oppermann12}, the foreground $\mathrm{RM_{fg}}$ around the position of M83
is $-34 \pm 10 \radm$. Since this value is an average over a much larger area,
the difference could be due, for example, to Faraday rotation in the halo of M83 or to local variations
in $\mathrm{RM_{fg}}$ of the Milky Way.

\begin{table}
\caption{ATCA configurations used to obtain the $\lambda13\cm$ radio continuum data of M83}
\begin{tabular}{cccc}
\noalign{\hrule\medskip}
Frequency &$\lambda$  &Configuration  &Date of \\
$[$MHz]   &[cm]       &               &observation\\
\noalign{\medskip\hrule\medskip}
2368 & 12.7   &750 A  &28/01/93 \\
2368 & 12.7   &750 B  &08/06/93 \\
2240 & 13.4   &750 B  &08/06/93 \\
2368 & 12.7   &750 C  &08/09/93 \\
2378 & 12.6   &750 C  &08/09/93 \\
\noalign{\medskip\hrule}
\end{tabular}
\label{arrays}
\end{table}

\subsection{APEX sub-mm observations of cold dust} \label{apex}

The $870\,\mu$m data were obtained with the Large APEX Bolometer Camera (LABOCA)
\citep{Siringo2009}, a 295-pixel bolometer array for continuum observations, operated at the
Atacama Pathfinder Experiment 12\,m-diameter telescope (APEX) \footnote{APEX is a collaboration
between the Max-Planck-Institut f\"ur Radioastronomie, the European Southern Observatory, and the
Onsala Space Observatory.} \citep{Gusten2006} at Chanjantor, Chile. We observed M83 in June, August,
and September 2008 in excellent weather conditions (the precipitable water vapour content ranged
from 0.1\,mm to 0.4\,mm).

M83 was mapped in the spiral raster mode (in the five-of-the-dice pattern), providing a fully
sampled map of the size $25\arcmin\times25\arcmin$ (compared to the LABOCA field of view of
$11\arcmin\times11\arcmin$) in each scan. The total on-source integration time was about
11.5~hours. The data were calibrated by observing Mars and Uranus together with the secondary
calibrator J1246-258, and the flux scale was found to be accurate within 15\%. The data were reduced
using the BOA (BOlometer array Analysis) software \citep{Siringo2009,Schuller2009}. After flagging
out bad and noisy pixels, the data were de-spiked, and correlated noise was removed for each scan.
Then the 280 scans were combined (weighted by the squared rms noise) to produce the final map shown
in the cold dust panel of Fig.~\ref{fig:maps}.

\subsection{Other data}\label{TAD}

In this paper the following data were used, referred to in Fig.~\ref{fig:maps} as indicated below:
\begin{itemize}
\item \textit{H\,{\sc{ii}}} (H$\alpha$), ionized hydrogen: 3.9\,m telescope of the
Anglo-Australian Observatory (AAO) with the TAURUS II focal reducer, observed on 20 May 1990
(S.~Ryder, priv.\ comm.);
\item \textit{H\,{\sc{i}}}, neutral hydrogen: VLA \citep{Tilanus93};
\item \textit{CO}, molecular hydrogen (traced by CO): SEST \citep{Lundgren04};
\item \textit{Gas}, total neutral gas ($\HI$ + CO) \citep{Lundgren04};
\item \textit{Warm dust}, infrared emission of warm dust at 12--18\,$\mu$m: ISO \citep{Vogler05};
\item \textit{Cold dust}, sub-mm emission of cold dust at 870\,$\mu$m (Section~\ref{apex});
\item \textit{I6}, total radio continuum intensity at $\lambda6\cm$, combined from VLA and Effelsberg
maps in total intensity \citep[see][for details]{Vogler05};
\item \textit{P6} and \textit{P13}, linearly polarized radio continuum intensity at $\lambda6\cm$ and
$\lambda13\cm$ (Figs.~\ref{fig:pi6} and \ref{fig:pi13}).
\end{itemize}

The infrared (warm dust) and $\HII$ maps provide a proxy to the star-formation rate in the
starburst regions (mainly in the bar region) and elsewhere.
Total radio continuum intensity has two components, a nonthermal (synchrotron) component, a tracer
of the components of the total (turbulent + ordered) magnetic fields in the sky plane, and a thermal
component, contributing about 12\% in M83 at $\lambda6\cm$ \citep{Neininger91}, which is neglected in
this paper.

Linearly polarized radio continuum intensity has a purely synchrotron origin. It is a tracer of
ordered magnetic fields with a preferred orientation within the telescope beam if Faraday depolarization
is small, which is generally the case in M83 at $\lambda6\cm$ and also in the outer disks of M83 at
$\lambda13\cm$.

The orientations of polarization ``vectors'' are ambiguous by multiples of 180\degr. As a
consequence, the ordered magnetic fields as traced by linearly polarized emission can be either ``regular''
fields, preserving their direction on large scales, ``anisotropic turbulent'', or
``anisotropic tangled'' fields with multiple field reversals within the telescope beam. To distinguish
between these fundamentally different types of magnetic fields observationally, additional Faraday rotation
data is needed. Faraday rotation is only sensitive to the component of the regular field along
the line of sight. Owing to the irregular distribution of Faraday rotation measures in M83 (Fig.~\ref{fig:rm}),
we cannot distinguish between the components of ordered fields.

The maps of total emission at $\lambda6\cm$ and $\lambda13\cm$ (I6 and I13), representing the total
magnetic field, are very similar, so that the $\lambda13\cm$ map was not
used for our analysis. The thermal contribution to the total radio emission of spiral galaxies at
these wavelengths is generally less than 20\% \citep{taba07}. The degree of polarization in M83 is
largely below 20\% at $\lambda6\cm$ \citep{Neininger93} and even lower at $\lambda13\cm$
(Fig.~\ref{fig:pi13}). The total radio emission at both wavelengths is thus dominated by
unpolarized synchrotron emission from turbulent magnetic fields.

On the other hand, the maps of polarized emission at $\lambda6\cm$ and $\lambda13\cm$ (P6 and P13),
representing the ordered magnetic field, are different. The signal-to-noise ratio is higher at
$\lambda13\cm$ because of the steep synchrotron spectrum, but Faraday depolarization is also stronger
at this wavelength. As a result, the polarized emission at $\lambda13\cm$ (Fig.~\ref{fig:pi13})
emerges mostly from the outer regions of the galaxy, while the lower sensitivity at $\lambda6\cm$
(Fig.~\ref{fig:pi6}) restricts the detected polarized emission to the inner regions. The sum
$1.91\,\textrm{P6} + \textrm{P1}3$ (P6+P13 in the following, see Fig.~\ref{fig:maps}), where P6 is
scaled by a factor of 1.91 to account for the average synchrotron spectral index of $-0.9$
\citep{Neininger93}, is used to represent the polarized emission from the whole disk of M83.
This procedure gives higher weights to the regions at intermediate distances from the centre that are seen
in both polarization images, but this is not relevant to the purpose of this paper.

All the maps were smoothed to a common resolution of $12\arcsec$, corresponding to $0.52\kpc$ at
the assumed distance to M83, except for the CO map that has a resolution of $23\arcsec$. As a
consequence, the map of total neutral gas also has a resolution of $23\arcsec$. All maps and their
wavelet transforms presented here are in the sky plane: the galaxy is oriented nearly face-on,
making a negligible correction for its inclination to the line of sight \citep[$24^\circ$
--][]{Tilanus93}.

\section{The method}
\label{sec:method}

In image analysis, wavelet-based methods are used to decompose a map into a hierarchy of structures
on different scales. Wavelets are a tool for data analysis based on self-similar basis functions
that are localized well in both the physical and wave-number domains. The localization of the
basis functions in the physical space distinguishes the wavelet transform from the otherwise
conceptually similar Fourier transform. One-dimensional and isotropic multi-dimensional wavelet
transforms are based on the space-scale decomposition of the data. (In other words, the family of
wavelets has two parameters, the location and the scale of the basis function.) Using the
continuous isotropic wavelet transform, a 2D image is decomposed into a 3D cube of wavelet
coefficients (obtained from the continuous wavelet transform by sampling it at a discrete set of
scales) with the scale as the additional, third dimension. Cross-sections of the cube are slices
that contain the image details on a fixed scale.  As a result, the wavelet transform preserves the
local properties of the image on all scales. If required, the original image can be synthesized
from the cube by summing over all scales. (This procedure is called the inverse wavelet
transformation.)

An \emph{anisotropic\/} wavelet transform is the convolution of the image with a set of wavelets
having different locations, sizes, and \emph{orientations}. Such a family of basic functions is
generated by the translations, dilations, and rotations of the basic wavelet. Applying the
two-dimensional anisotropic wavelet transform to an image generates a four-dimensional data set that
represents a space-scale-orientation decomposition. Fixing the space and scale parameters --
based on some objective criteria -- enables one to track the orientation of an elongated structure.
An extended description of the continuum wavelet transform can be found in various books, for
example, in \citet{Holschneider95}. In extragalactic radio astronomy, galactic images have been
analysed using isotropic wavelets by \citet{frick01,taba07,taba13} and anisotropic wavelets by
\citet{patrikeyev06}.

Both isotropic and anisotropic 2D wavelets can be constructed from a popular real-valued wavelet,
the Mexican Hat (MH). In 1D, the MH is given by $\psi(x)=(1-x^2)\exp(-x^2/2)$. The MH can be
generalized to 2D, leading to an isotropic basis function,
\begin{equation}
\psi(r)=(2-r^2)\exp\left(-{{r^2}/{2}}\right),
   \label{eq:mh}
\end{equation}
where $r = (x^2 + y^2)^{1/2}$. Another possibility is an anisotropic wavelet obtained by
supplementing the 1D MH in one dimension with a Gaussian-shaped window along the other axis. The
result is an anisotropic wavelet introduced by \cite{patrikeyev06}, called the Texan Hat (TH):
\begin{equation}
   \psi(x,y) = (1-y^2)\exp\left( -\frac{x^2+y^2}{2}   \right).
   \label{eq:th}
   \end{equation}
This wavelet is sensitive to the structures elongated along the local $x$-axis. Rotation in the
$(x,y)$-plane, $x\to(x\cos\varphi+y\sin\varphi),$ and $y\to(y\cos\varphi-x\sin\varphi)$  in
Eq.~(\ref{eq:th}), introduces the angle $\varphi$ counted from the $x$-axis in the counterclockwise
direction as the wavelet parameter.  Finally, a set of basis functions of different sizes and
orientations is obtained by applying both dilation and rotation to the basic wavelet. Since the TH
is a symmetric with respect to rotation by 180\degr, $\psi_{\varphi}(x,y) = \psi_{\varphi +
\pi}(x,y)$, it is sensitive to the \emph{orientation} of an elongated structure but not to its
\emph{direction,\/} which could be defined for structures in vector fields, such as magnetic field or
velocity. The anisotropy of the TH makes it especially convenient in analyses of galactic spiral
patterns. A detailed description of the technique and the illustration of how it works with test
images and the spiral structure maps of the galaxy M51 were given by \citet{patrikeyev06}.

The continuous wavelet transform of a 2D map $f(\vec x)$, $\vec{x}=(x,y)$, is defined by
\begin{equation}
W(a,\varphi,\vec x) = \frac{1}{a^\kappa} \iint_{R} f(\vec x') \psi_{\varphi} \left(
\frac{\vec x'-\vec x}{a}\right) \,\dd^2 \vec x'\,, \label{cwt}
\end{equation}
where the integration is extended over the image area $R$, and the normalization factor $a^{-\kappa}$
allows one to fine-tune the interpretation of the wavelet transform: $\kappa=3/2$ is for direct
comparison of the wavelet spectra to the Fourier spectra, whereas $\kappa = 2$, used here, ensures
that a power-law approximation to the wavelet spectrum has the same exponent as obtained from
the second-order structure function  \citep{frick01}. In the case of an isotropic wavelet,
$\varphi$ should be omitted in Eq.~(\ref{cwt}).

The maximum value of the wavelet transform $W(a,\varphi,\vec x)$ over all position angles $\varphi$
for a given scale $a$,
\begin{equation}
W_\mathrm{m}(a,\vec x) = \max_{0\leq\varphi\leq\pi}{W(a,\varphi,\vec x)}\,, \label{wtmax}
\end{equation}
can be used to quantify the anisotropic fraction of structures identifiable in the image. The
orientation of the anisotropic structure is then given by the corresponding value of the position
angle, $\varphi_\mathrm{m}$, such that $W_\mathrm{m}= W(a,\varphi_\mathrm{m},\vec x)$.

The distribution of the energy content of the signal, $f^2$, on the scales and at the orientations can
be characterized by the {\it wavelet power spectrum\/}, defined as the energy density of the
wavelet transform on scale $a$ and at orientation $\varphi$ integrated over the image,
\begin{equation}
M(a,\varphi) = \iint_R |W(a,\varphi,\vec x)|^2 \,\dd^2 \vec x \, . \label{w_spec}
\end{equation}
The isotropic wavelet spectrum is obtained as
\begin{equation}
M(a) = \iint_R |W(a,\vec x)|^2 \,\dd^2 \vec x \, ,  \label{w_spec_1}
\end{equation}
where the wavelet coefficients $W(a,\vec x)$ are obtained using the isotropic wavelet
(\ref{eq:mh}).

\section{Results}

We applied the MH and TH wavelet transforms to the maps of M83 presented in Section~\ref{TD} with
the aim of identifying structures (and their orientation, if appropriate) revealed in various tracers
with particular emphasis on the spiral structure.

\subsection{Isotropic wavelets: spectra and correlations}

\begin{figure}
\centerline{\includegraphics[width=0.45\textwidth]{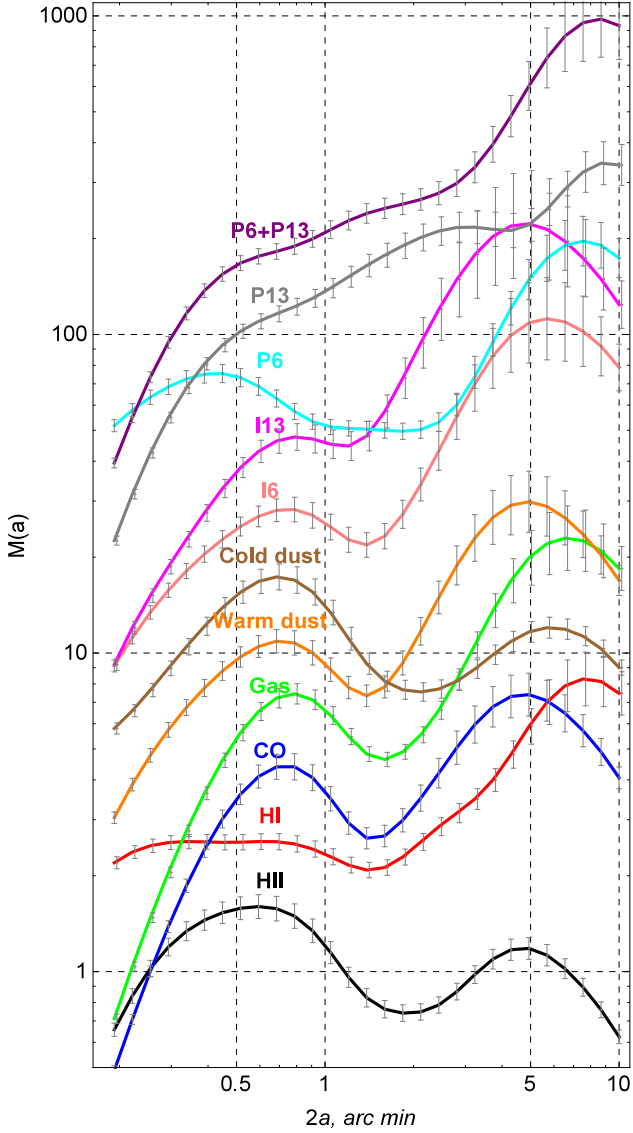}}
\caption{Wavelet power spectra of the M83 images, obtained with the isotropic 2D Mexican Hat
from bottom to top: $\HII$, $\HI$, CO, total neutral gas, warm dust, cold dust, total synchrotron
intensity at $\lambda6\cm$, polarized intensity at $\lambda6\cm$, polarized intensity at
$\lambda13\cm$, and a weighted sum of the polarized intensities at $\lambda6\cm$ and $\lambda13\cm$
defined in Section~\ref{TAD}. All the spectra have been multiplied by various factors chosen to
avoid overcrowding of the curves: what matters is the relative distribution of the wavelet
spectral power between scales in a given image rather than the relative power in different images.
Error bars are evaluated from the variance of the sample mean.  }
\label{spectra_new2}
\end{figure}

We start our analysis with the wavelet spectra $M(a)$ calculated for each tracer using the
isotropic MH, Eq.~(\ref{eq:mh}). Our aim here is to compare the dominant spatial scales of the
different maps, i.e., the scales at which the wavelet power spectra have a maximum. The magnitudes
of the maxima are not informative because they depend on the signal intensities measured in different
units in some maps. To present the results in one plot, we multiplied each spectrum by a
factor chosen to avoid excessive overlapping of the curves. The isotropic MH power spectra are
shown in Fig.~\ref{spectra_new2}.

\begin{figure}
\includegraphics[width=0.43\textwidth]{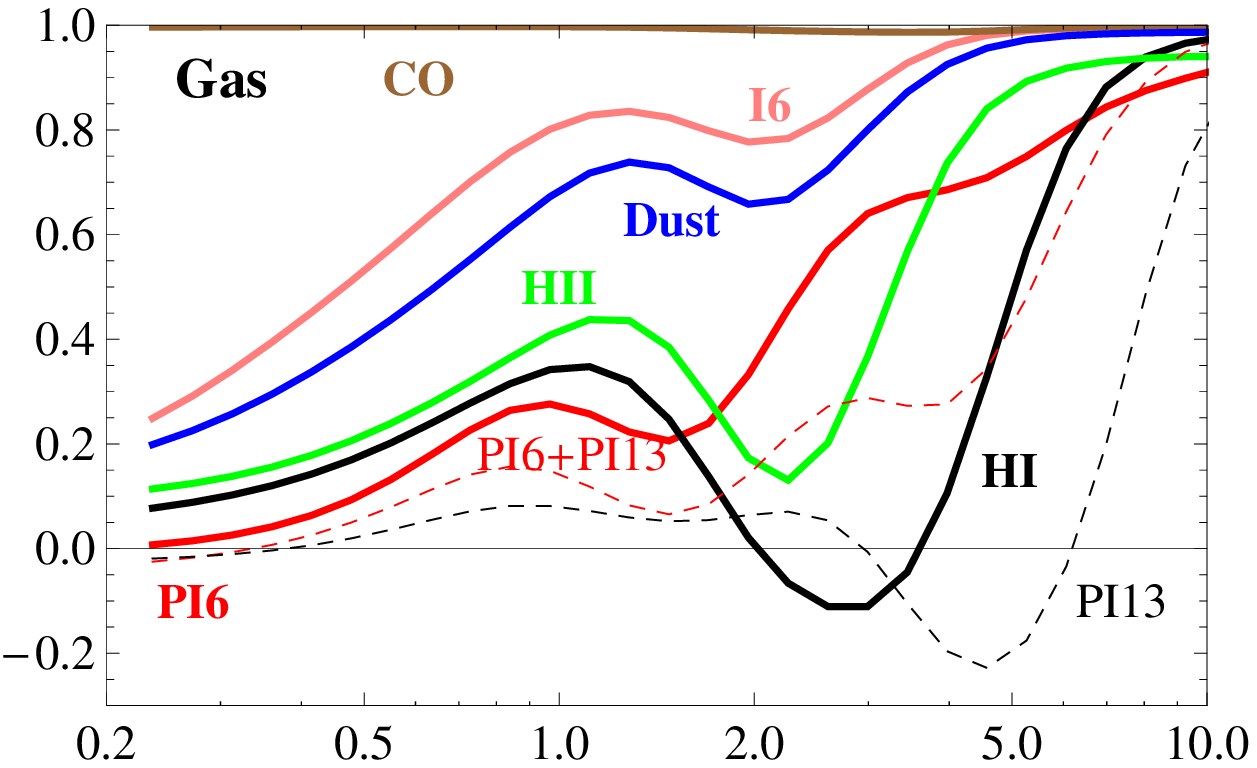}
\includegraphics[width=0.43\textwidth]{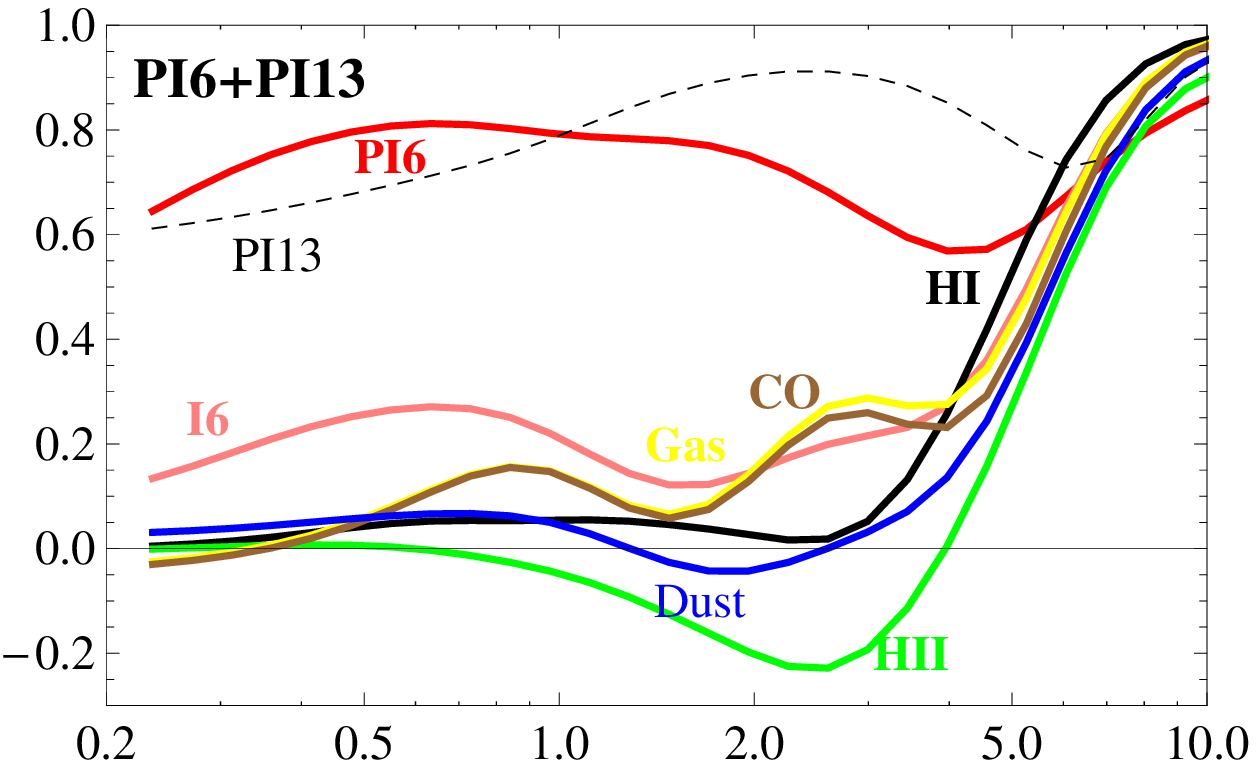}
\caption[]{Wavelet cross-correlations between the M83 maps as a function of scale (in arcmin).
Correlations between the total neutral gas and other tracers ($\HII$, $\HI$, warm dust, I6, P6+P13)
are shown in the upper panel and correlations of the polarized emission with other tracers in the
lower panel.}
 \label{wav_cor_new}
\end{figure}

The spectra reveal that the distribution of the ISM seen in
various tracers is characterized by
certain dominant scales.\footnote{The scale of a structure used here, $2a$, represents the diameter
rather than the radius of the structure.} The maximum at the largest scales corresponds to the size
of the galaxy as a whole, $2a \approx 10\arcmin$ in both polarized emission (either $\lambda6\cm$
or $\lambda13\cm$) and $\HI$. The images of I6, dust, total neutral gas, and $\HII$ are more compact,
and the corresponding spectral maximum appears on smaller scales, $2a\approx
5\arcmin\textrm{--}7\arcmin$. The decrease in the spectra on the largest scales occurs
due to the large void areas in the signal maps.

\begin{figure}
\centering
\includegraphics[width=0.35\textwidth]{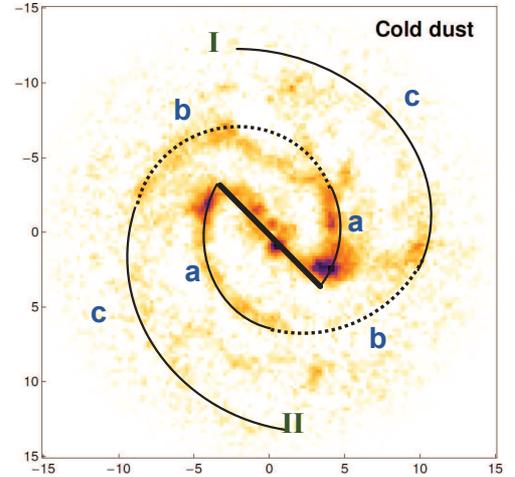}
\caption{\label{fig-arms}Schematic representation of the spiral arms and bar (thick solid line)
in M83. Each of the two arms (labelled I and II) is separated into three sections, labelled as a,
b, and c and shown as solid, dotted, and solid lines, respectively. The background image represents the cold dust
distribution. The axis scale is in kpc, assuming the distance to M83 of 8.9\,Mpc.}
\end{figure}

Maxima on smaller scales are produced by structures inside the galactic image, bar, and spiral arms.
A maximum on the scale $2a \approx 0.7\arcmin$ is prominent in I6, dust, CO, and total neutral gas.
The spectrum of $\HII$ also has a local maximum on this scale, but it is much broader and extends
down to $2a\approx 0.4\arcmin$. The second maximum of the polarized intensity spectrum at
$\lambda6\cm$ (P6) is at $2a = 0.4\arcmin$, probably reflecting sharp features in the bar.
Remarkably, the spectrum of the other magnetic field tracer, the polarized emission at
$\lambda13\cm$ (P13), is quite different when it grows with $a$ reaching a weak maximum at $2a
\approx 3 \arcmin$. The P13 map contains large structures that are absent in the other maps. This scale can
be related to the outer, broad spiral arms. The $\HI$ spectrum is nearly flat on small scales with
a weak, broad maximum at $0.2\arcmin < 2a < 1\arcmin$. The values of the peak scales should be
considered with an uncertainty of about 20\% because of the limited scale resolution of the wavelet.

The wavelet cross-correlation between two maps (1 and 2) is defined as \citep[see][]{frick01}
\begin{equation}
r_w(a) = \frac{\iint_R W_1(a,\vec x) W_2(a,\vec x) \,\dd^2 \vec x}{[M_1(a)M_2(a)]^{1/2}} \,,
\label{cor_w}
\end{equation}
where the integration is carried over the map area. This quantity, designed to characterize
scale-by-scale correlations in Maps 1 and 2, is sensitive to structures in the two maps that
have similar scales and a similar position, but may not overlap as do, for example, spiral arm segments with a
relative shift in position. The standard cross-correlation function fails to detect such subtle
correlations, whereas examples presented by \citet{frick01} demonstrate the efficiency of the
wavelet cross-correlation.

The wavelet correlations have been calculated for all pairs of tracers in the whole scale range
available. Figure~\ref{wav_cor_new} shows the cross-correlations of the total neutral gas
and all other tracers in the upper panel, whereas the lower
panel is for the scale-wise cross-correlation between the polarized emission P6+P13 and all other
tracers.

The tracers of matter (gas and dust) are more or less well correlated with each other. Not
surprisingly, since they are not independent, this is true at all scales for the CO and the total
neutral gas. Dust, $\HII$ and $\HI$ have maximum correlation with the total neutral gas near the
scale $0.7\textrm{--}0.9\arcmin$ (1.8--2.3\,kpc). The correlation coefficient for dust and gas is
about $0.7,$ while the correlation coefficient of $\HI$ and the total gas is below $0.4$). In a
striking contrast, the ordered magnetic field tracers are almost uncorrelated with the total
neutral gas and dust on scales up to $2\arcmin$ and even exhibit some anticorrelation (e.g. gas
versus P13 on the scale of about $2\arcmin$ and P6+P13 versus $\HII$ on the scale of about
$1.5\arcmin$).

\begin{figure}
\centering
\includegraphics[width=0.35\textwidth]{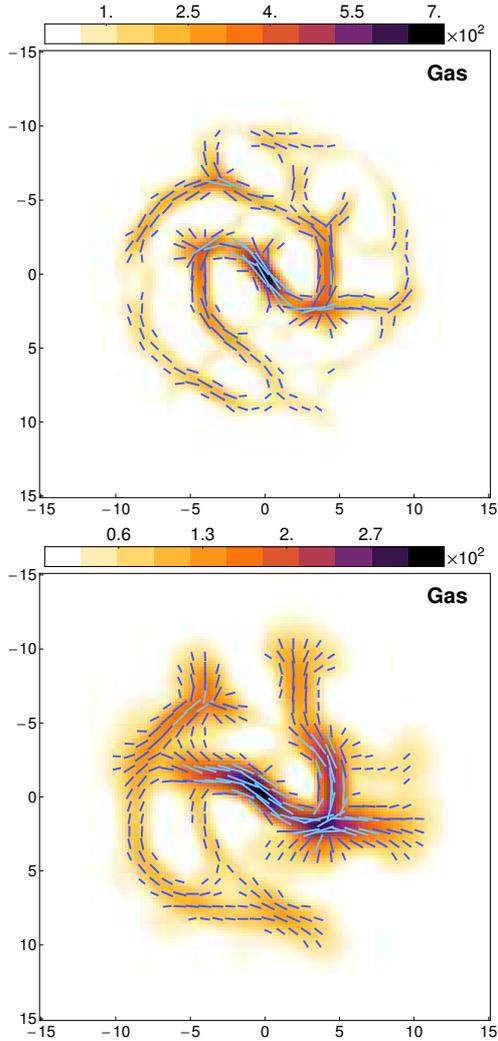}
\caption{\label{fig:cmap4g}Anisotropic wavelet transforms of the total neutral gas on the scales
$2a= 0.7\arcmin$ (top) and $1.4\arcmin$ (bottom). The wavelet orientation angles
are shown by bars whose lengths are proportional to the magnitudes of the wavelet transform. The
axis scale is in kpc.}
\end{figure}
\begin{figure}
\centering
\includegraphics[width=0.35\textwidth]{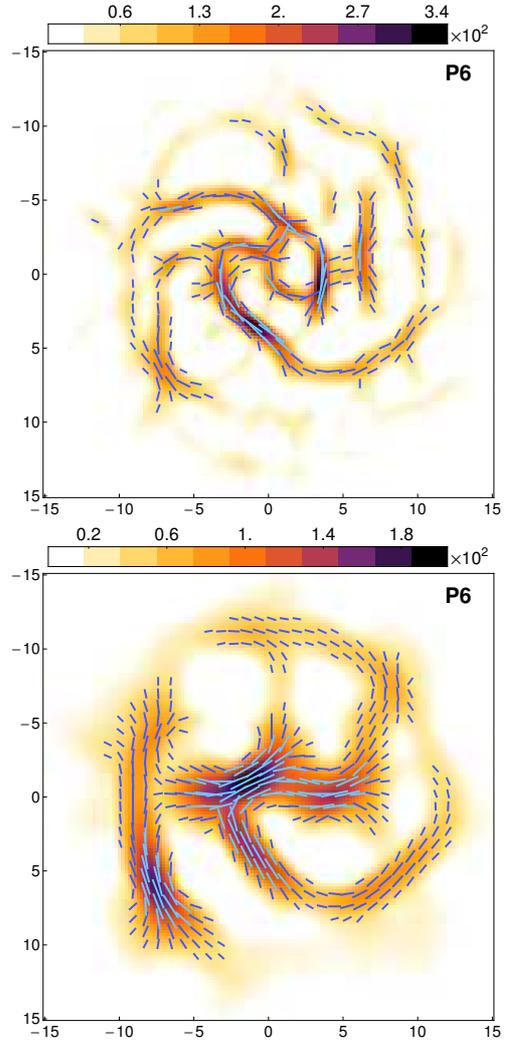}
\caption{\label{fig:cmap4p}As in Fig.~\ref{fig:cmap4g} but for P6, the polarized radio emission at
$\lambda6\cm$.}
\end{figure}

The correlations between particular tracers can be summarized as follows. (We do not show all the
plots.) The correlation between the total radio continuum (I6) and infrared (warm dust) intensities
is high on all scales, similar to the situation in NGC\,6946 \citep{frick01,tabatabaei2013a} and
M33 \citep{taba13}; this can be explained by the contributions of thermal radio emission and
synchrotron emission from magnetic fields that are closely related to molecular gas
\citep{niklas97}. The correlation between the radio synchrotron and infrared intensities is known
to be worse on small scales due to the propagation of cosmic-ray electrons \citep{taba13}, so that
the correlation coefficient is expected to be below 0.5 on scales that are smaller than the cosmic ray
diffusion scale along the large-scale magnetic field of about $0.2\arcmin$ or $0.5\kpc$. The ratio
of the ordered and turbulent field strengths $q\approx0.5$ follows from the average degree of
synchrotron polarization of $p\approx20\%$ in the M83 disk at $\lambda6\cm$ using the relation
between $p$ and $q$ for a uniform cosmic-ray distribution \citep[Eq.~(2) in][]{B07} (under perfect
equipartition between cosmic rays and magnetic field, $q\approx0.43$). With the former value of
$q$, M83 fits the relation between the parallel diffusion scale and the degree of field ordering
found by \citet{taba13}). A more detailed analysis would need a separation of synchrotron and
thermal radio emission, which is beyond the scope of this paper.

The correlation between $\HII$ and infrared (warm dust) intensities is also high on all scales,
except for a minimum around $2\arcmin$, probably because of the bar that is bright in the infrared but
weak in $\HII$.

The I6, dust, and $\HII$ maps display high correlations on small scales ($r_w>0.5$ for $0.2\arcmin <
2a < 1 \arcmin$), where correlations of the other tracers are weak. The maps of polarized intensity
P6 and P13 display quite different properties. The P6 map shows little correlation with most tracers on all
scales, except the largest ones, and an anticorrelation with $\HI$ on the scale $2a \approx
2.5\arcmin$. All correlations with P13 show an absence of correlation on small scales and have a
deep minimum (anticorrelation) on the scale of about $3\arcmin$ with $\HII$, dust, CO, and I6 (but
not with $\HI$). We note that P13 and $\HI$ are the two maps with emission distributed over the largest
areas, but just this pair of maps gives the minimum correlation over the whole range $0.2\arcmin <
2a < 4\arcmin$ (neither correlation nor anticorrelation). The highest anticorrelation ($r_w\approx
-0.4$ is found between $\HII$ and P13 at $2a\approx 3\arcmin$. The $\HII$ and P6 have a weaker
anticorrelation, $r\approx -0.2$ at $2a\approx 1.7\arcmin$. The anticorrelation with $\HII$ around
this scale indicates a general shift between the optical and magnetic spiral arms, similar to the shift
in NGC\,6946 \citep{frick01}.

On scales of $\le 0.7\arcmin$, corresponding to the maximum in the spectrum of P6, the emission
emerges from the inner part of the galaxy. In P13, however, the emission from the central part of
the galaxy is strongly suppressed by Faraday depolarization, and the larger scales attributed to the
outer parts of the galaxy dominate in the wavelet spectrum. Thus the anticorrelation with P13 is
probably the effect of depolarization.

\subsection{Material versus magnetic patterns}

A more detailed comparison of the spatial patterns in the interstellar matter and magnetic fields
is facilitated by the use of the anisotropic wavelet introduced in Section~\ref{sec:method}. It is
convenient to introduce a simple reference pattern of the dominant material structures shown in
Fig.~\ref{fig-arms} that includes two spiral arms, labelled Arms~I and~II, emerging from the
opposite ends of the bar shown by a thick straight line in the figure. Each arm is divided into
three segments, a, b, and c, because some of them are not visible in all the tracers considered
here.

\begin{figure*}
\centering
\includegraphics[width=0.9\textwidth]{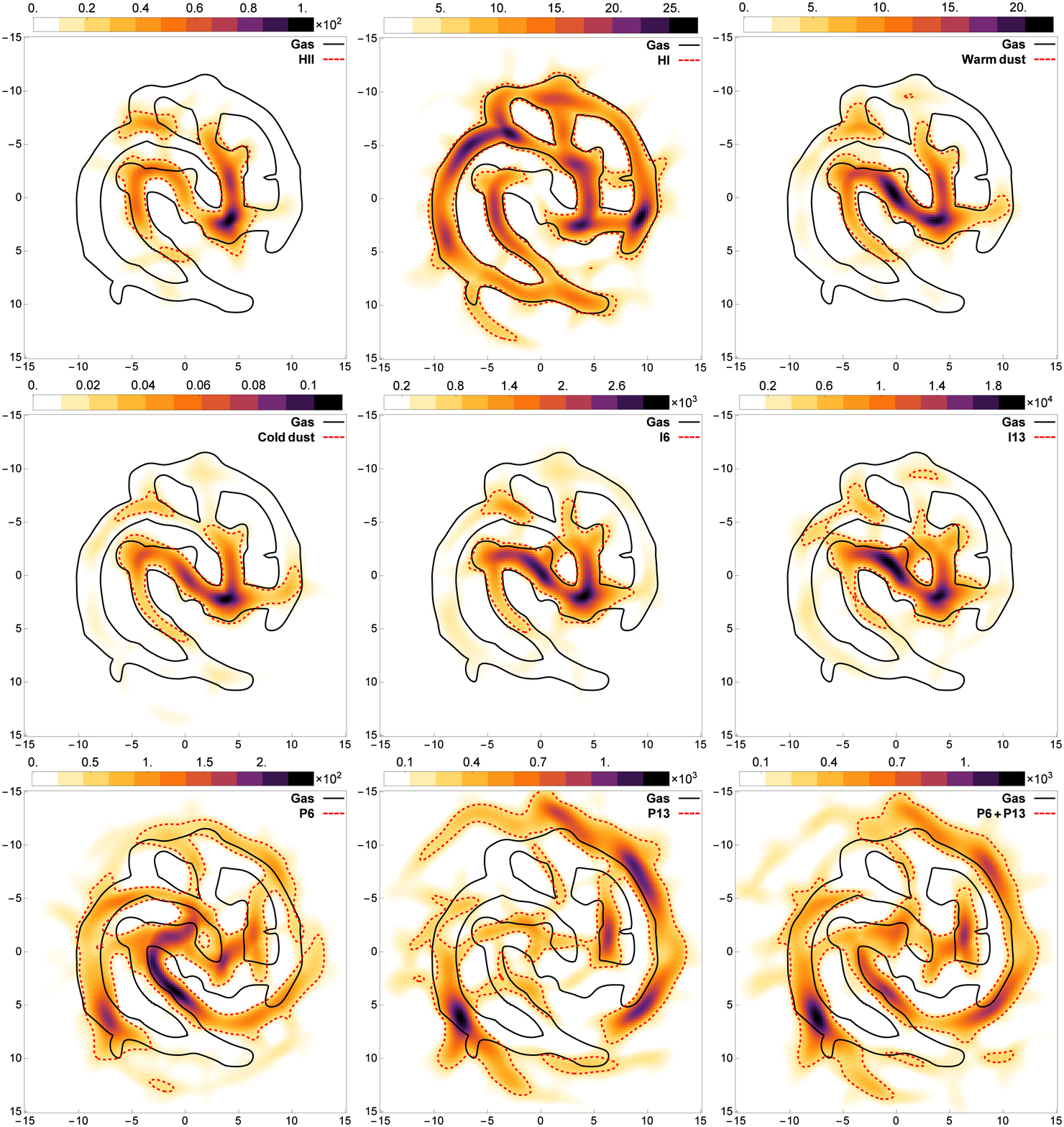}
\caption{\label{fig:gmap1}Anisotropic (TH) wavelet transforms of various tracer distributions
at $2a=1\arcmin$ are shown (dotted contour at 20\% of the maximum intensity), together with the similar
wavelet transform of the total neutral gas distribution (solid contour at 20\% of the maximum intensity).
The axis scale is in kpc.}
\end{figure*}

An anisotropic wavelet transform is useful on those scales where the spatial structures are clearly
elongated, i.e., on the scales comparable to and exceeding their thickness, $0.5\arcmin < 2a <
2\arcmin$ in the case of M83. Figure~\ref{fig:cmap4g} shows the anisotropic wavelet transform of
the total neutral gas distribution obtained with the TH wavelet. A similar map for the polarized
intensity P6 is shown in Fig.~\ref{fig:cmap4p}. We selected two representative scales to show in
these figures, $2a\approx 0.7\arcmin$ and $1.4\arcmin$ as suggested by the power spectra of
Fig.~\ref{spectra_new2}. The maximum value $W_\mathrm{m}$ of the wavelet transform over all
position angles $\varphi$ on a given scale is shown as colour-coded, whereas the corresponding wavelet
orientation angle $\varphi_\mathrm{m}$ is indicated with a bar.

Figures~\ref{fig:cmap4g} and \ref{fig:cmap4p} demonstrate clearly that the distribution of
polarized radio emission is more structured than that of the template pattern (Fig.~\ref{fig-arms}).
On the smaller scale, $0.7\arcmin$, the bar is prominent in the total neutral gas, whereas the polarized
structures are offset from the bar axis and are hardly aligned with it. Their overall pattern is
similar to the barred galaxy NGC\,1097, where it suggests the amplification of
magnetic fields by compression and shearing gas flow in the dust lanes displaced from the bar axis
\citep{Becketal05}.

\begin{figure*}
\centering
\includegraphics[width=0.9\textwidth]{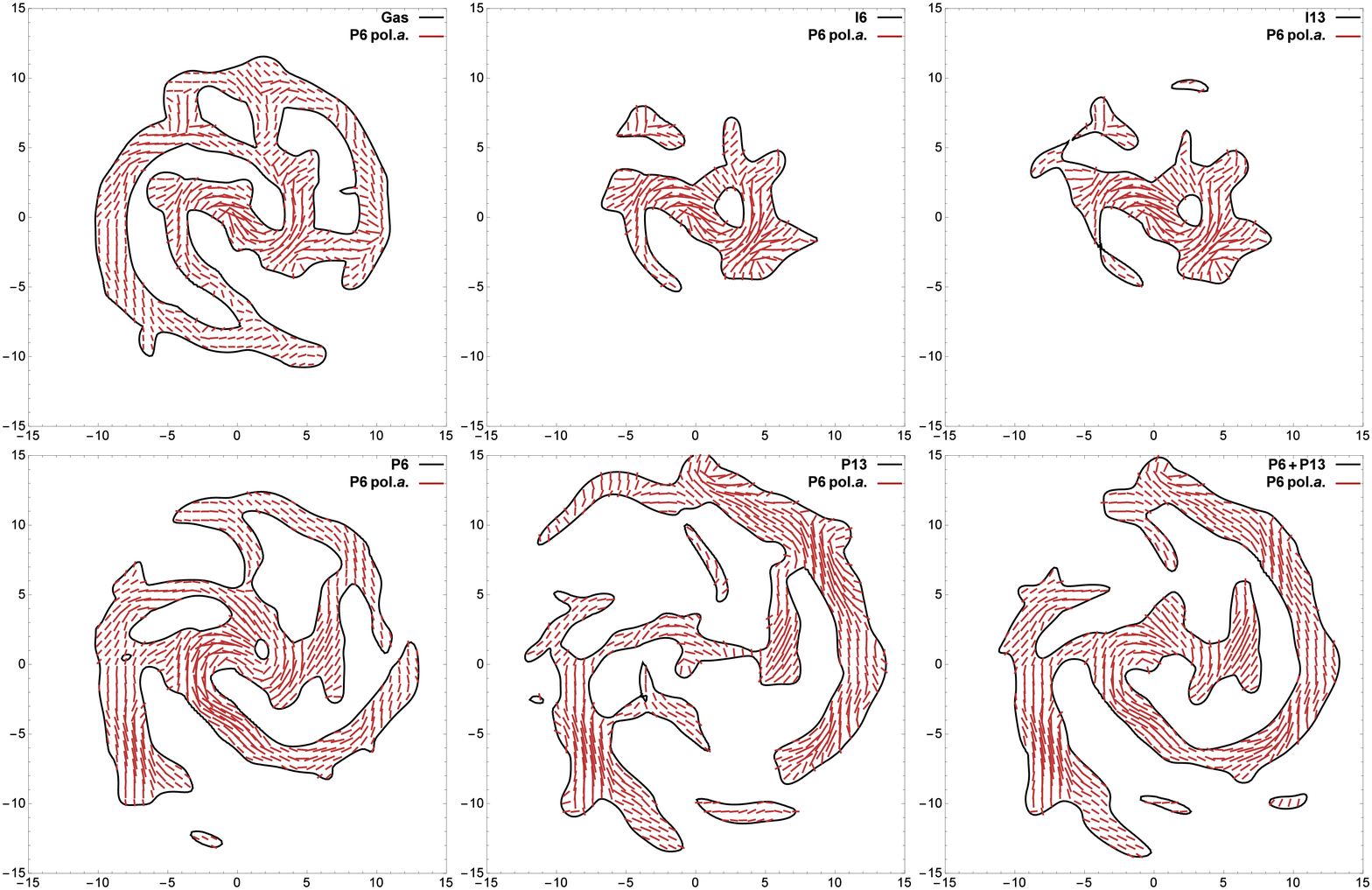}
\caption{\label{fig:pmap1}Elongated structures on the scale $2a=1\arcmin$ of the anisotropic
wavelet transform in the tracer indicated in the upper right corner of each frame are shown with
the solid contour drawn at 20\% of the maximum intensity in that tracer in the arm region. Red
dashes, representing the $B$ vectors of the polarized radio emission at $\lambda6\cm$, with length
proportional to the polarized intensity, are shown in the same regions. }
\end{figure*}

On the larger scale, $1.4\arcmin$, most features visible on the smaller scale remain, although some
smaller structures of the upper panel have merged on this larger scale. The P6 map has a
well-pronounced long arm extended from west to north in the outer part of the image (Part~c of
Arm~I) apparently merging with Arm~II in the east, but this appears to be an artefact of stronger,
positively biased noise in polarized emission at large distances from the map centre that affects
the wavelet transform.

The patterns isolated with the anisotropic TH wavelet in the material tracers are compared with
those in the ordered magnetic field in Fig.~\ref{fig:gmap1} where the total neutral gas density is
chosen as the reference variable.  In all the panels, the solid line corresponds to 20\% of the
maximum intensity in the arm region. \footnote{We do not use the absolute maximum because it is
located in the bright core that dominates the gas map.} Arm II is very visible in all three
segments (IIa, IIb, and IIc) in the gas distribution. The CO distribution on this scale is very
similar to that of the total neutral gas (unsurprisingly, because molecular gas dominates over
$\HI$ in the inner part of the galaxy) and is not shown here. The only difference of the CO map is
that the segments IIa, IIb, IIc are separated from each other by small gaps. The warm and cold dust
have the same pattern as the gas, with the exception of the segment IIc. The $\HII$ image contains
only the inner part of Arm II, while $\HI$ clearly displays the whole of Arm II. Concerning Arm I,
only the first segment Ia is visible in gas, CO, and warm and cold dust. Its outer part Ic is
prominent in the $\HI$ image but not in the other material tracers. To conclude, there are no
discernible differences in the appearance, in various material tracers, of the bar as well as of Arms
Ia, IIa, and IIb, whereas Arms Ib, Ic and IIc appear different in different material tracers.

The tracers of the total magnetic field, I6 and I13, are presumably dominated by small-scale,
random magnetic fields. The structures in I6 and I13 are very similar to those in $\HII$ and the
warm and cold dust, with the bar and Arms Ia and IIa being prominent. The spiral segments
Ib and Ic are absent in I6 and I13.

The patterns in P13 and the combined P6+P13 are similar to each other on this scale with prominent
outer spiral segments Ic and IIc, as well as Arm Ib. As we quantify below, the ridges in radio
polarization are noticeably different from the gas ridges in the bar, often being shifted with
respect to each other. Arm IIb is not visible in polarized radio emission, whereas Arm IIa is
shifted with respect to IIa of the total neutral gas. The segment Ib occurs in P6+P13, thus
delineating the whole Arm I. Arm IIc largely overlaps the corresponding gas arm in both P6 and P13.
Remarkably, the polarized Arm IIa  is displaced towards the outer galaxy with respect to the
corresponding gas arms by 2--3\kpc, whereas Arm Ia is shifted inwards by a comparable magnitude. The
mutual displacements of the gaseous and polarized arms are, on the one hand, systematic and
coherent over several kiloparsecs along the arms, but on the other hand, they take different
directions in different arms even in the same range of galacto-centric radii. Arm IIa
(south-east of the galaxy, the "Gas" and "P6+P13" panels) shows the clearest picture
of a polarized arm being parallel to the gaseous arm along about 8\,kpc. On the other hand,
in Arm IIc (the next outer arm in the south-east) a polarized arm overlaps the gaseous arm
nearly perfectly at least along 12\,kpc, and only its end
deviates outwards from the gaseous arm.  This implies that the mechanisms producing such a
displacement are more complex than just advection of a large-scale magnetic field from the gaseous
arms by the rotational velocity or the enhanced tangling of a large-scale magnetic field within the
gas arms. Furthermore,
Arms Ib and Ic appear mainly in the radio polarization data but not in the total radio intensity
(dominated by small-scale magnetic fields) and are visible in only one material tracer, $\HI$. The
physical nature of the magnetic Arms IIc, Ib, and Ic is likely to be different from Arms
I, IIa, and IIb (see Sect.~\ref{discussion}). A more detailed comparison of the gaseous and magnetic
spiral arms and their segments arms is presented in the next section.

\subsubsection{The orientations of the ordered magnetic field and the spiral arms}
\label{ORMFSA}

The wavelet representation allows us to investigate anisotropic structures on any given scale.
Figure~\ref{fig:pmap1} helps for appreciating the complicated relation between the orientation of magnetic vectors
and the orientation of structures in the distribution of the various tracers of spiral arms.
The wavelet coefficients show the intensity of the signal that is smoothed over a domain of
scale $a$. Thus, gaps in the original maps seen on smaller scales can be filled on larger scales.

The magnetic vectors (uncorrected for Faraday rotation) in the magnetic arms, shown in the lower left-hand panel
of Figure~\ref{fig:pmap1}, are well (and yet imperfectly) aligned with the arms of polarized emission,
while the magnetic vectors in the bar region are noticeably inclined to its axis, especially near the ends of
the bar.

\begin{figure}
\centering
\includegraphics[width=0.4\textwidth]{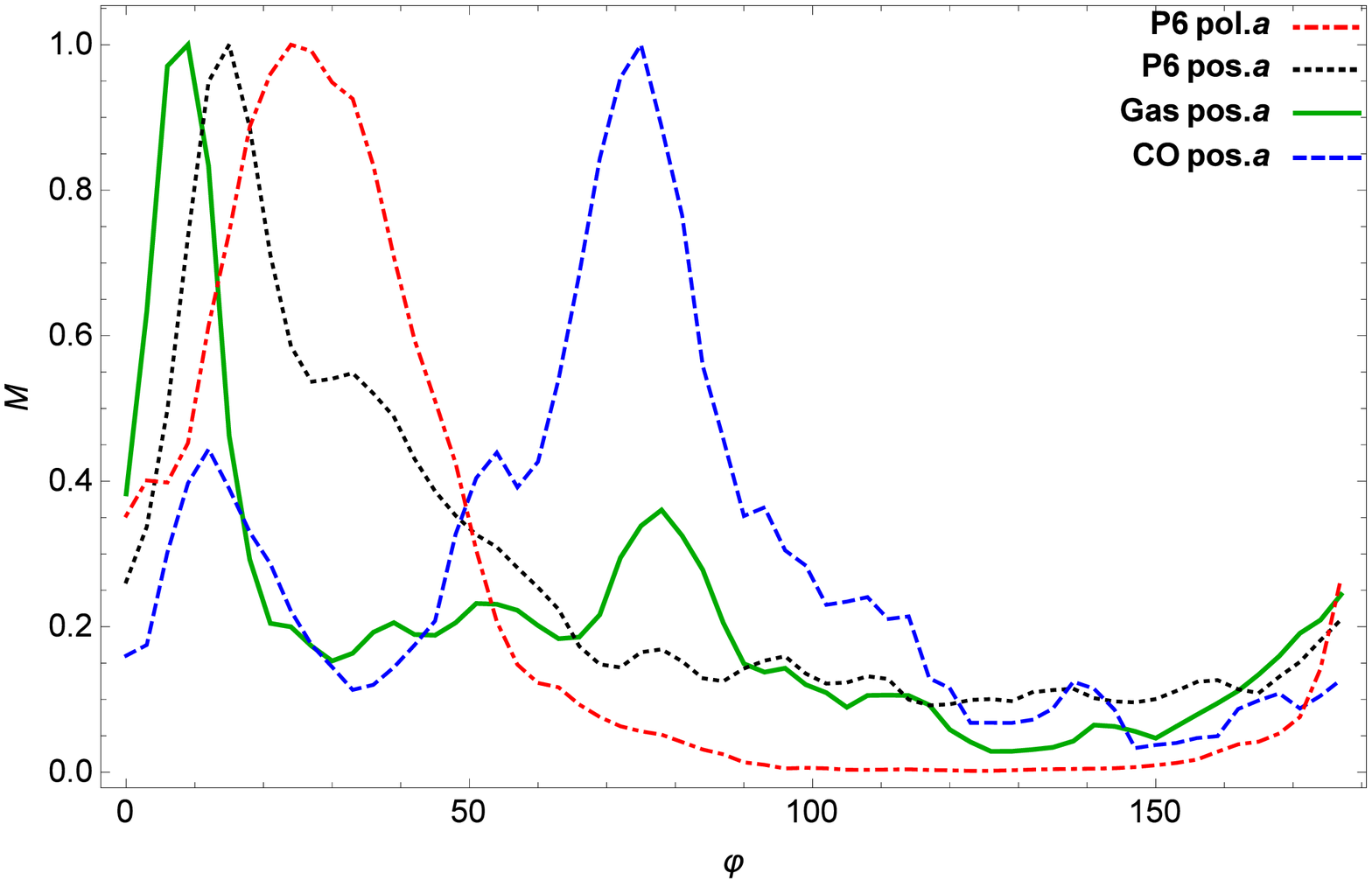}\\
\includegraphics[width=0.4\textwidth]{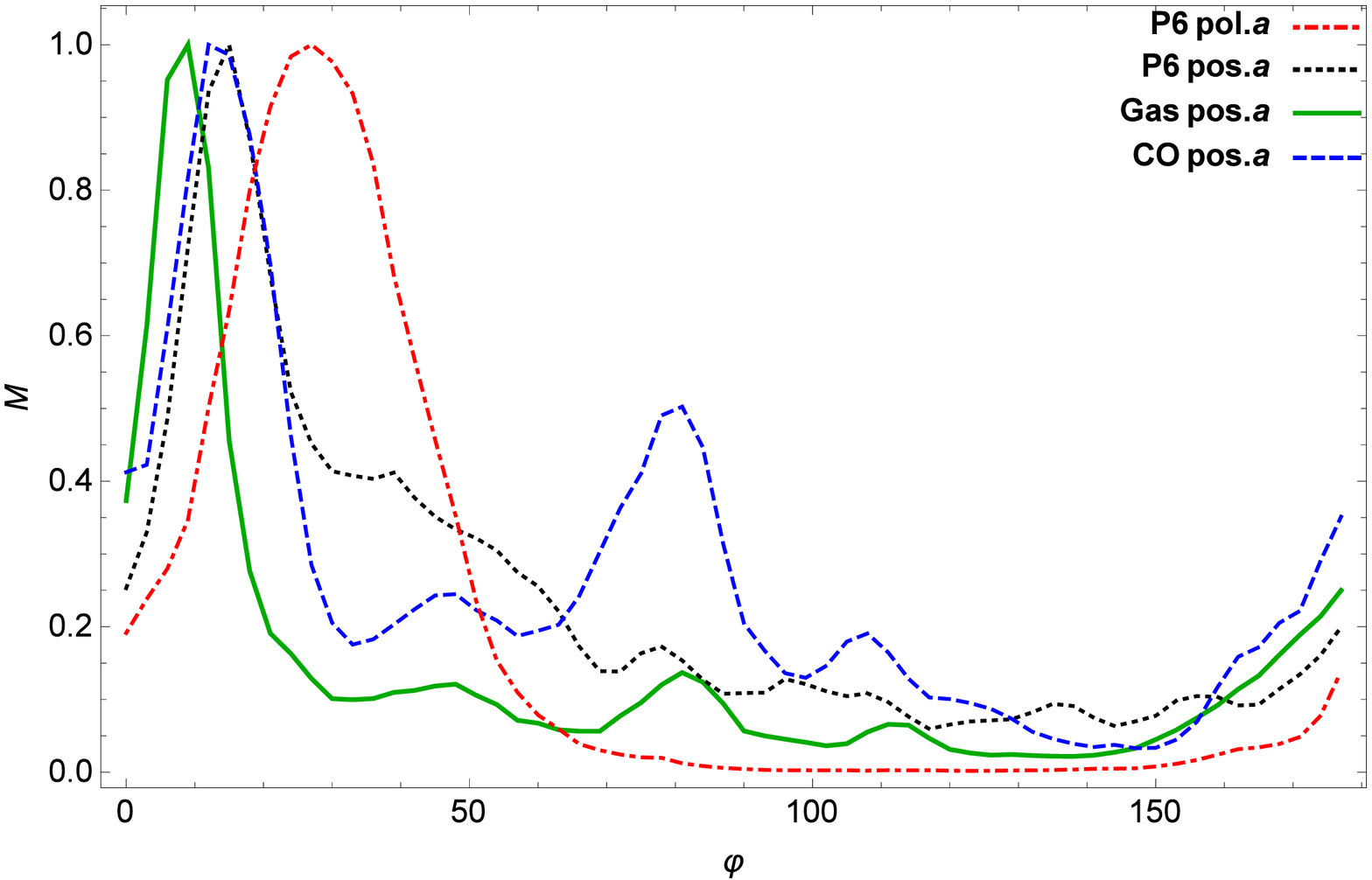}
\caption{\label{fig-angles} Angular wavelet power spectra of pitch angles $\varphi$
on the scale $2a=1\arcmin$ normalized to unit maximum: the pitch angles $\varphi_\mathrm{m}$ of the structures
of the polarized intensity P6 (black dotted), neutral gas (green), and CO (blue dashed).
The spectrum of the pitch angles $\varphi_\mathrm{p}$
of the magnetic field vectors defined in Eq.~(\ref{w_spec_p})
is shown with the red dot-dashed curve. The pitch angles are measured clockwise from the tangent
to the local circumference on the sky plane. The upper panel shows the spectra for the whole galaxy,
while the lower panel shows them for the region of spiral arms (i.e. with the bar region excluded).}
\end{figure}

The near-alignment of magnetic field with the spiral structures requires quantifying. For this
purpose, we compared the orientations of the magnetic field vectors (\textit{polarization angles})
with the orientations of the anisotropic (elongated) structures on any given scale.
Figure~\ref{fig-angles} shows the normalized angular power spectra $M(a,\varphi_\mathrm{m})$
of the pitch angles $\varphi_\mathrm{m}$ of the structures on a fixed scale
$2a=1\arcmin$ in the distributions of polarized intensity P6, labelled as "P6~pos.a", and of the
interstellar gas, i.e. the total neutral gas, "Gas~pos.a", and molecular gas, "CO~pos.a". Here,
$\varphi_\mathrm{m}$ is measured from the tangent to the local circumference in the plane of the
sky. \footnote{We neglect the small difference in the pitch angle measured in the plane of the galaxy
because the inclination of M83 to the line of sight is only $24^\circ$ \citep{Tilanus93}.}
The spectra of Fig.~\ref{fig-angles} characterize the orientation of the ridges in the distributions of
interstellar gas and polarized radio intensity on average over the corresponding region, the whole
galaxy in the upper panel, and the spiral arm region in the lower panel.

Figure~\ref{fig-angles} also presents the angular spectrum $M(\varphi_p)$ of the magnetic pitch
angles $\varphi_\mathrm{p}$ derived from the polarized intensity at $\lambda6\cm$.
This spectrum is defined as the integral of polarized intensities over all positions where the pitch
angle of the magnetic field vectors at $\lambda6\cm$, $\chi$, is equal to a given $\varphi_\mathrm{p}$:
\begin{equation}
M(\varphi_\mathrm{p}) = \iint_{pitch(\chi)=\varphi_\mathrm{p}} {\mathrm{P6}}(\vec x) \,\dd \vec x \,. \label{w_spec_p}
\end{equation}
The angle $\varphi_\mathrm{p}$ is also measured from the local circumference, thus representing the magnetic
pitch angle (uncorrected for Faraday rotation, which is small at $\lambda6\cm$, and for the
inclination of the galaxy to the line of sight, which is only modest).

The most pronounced difference between the spectra shown in the two panels of Fig.~\ref{fig-angles}
is in the position angles of the anisotropic CO structures dominated by the bar that provides a strong maximum
at pitch angles $\varphi_\mathrm{m}$ of about $70^\circ-80^\circ$ ($90^\circ$ would correspond to a structure
aligned along the radial direction). When the bar is excluded, this peak becomes twice lower than the
other CO maximum centred at about 13$^\circ$, with a half-width at half-maximum (HWHM) of $9^\circ$
that arises from the spiral arms. The sharp peak in the spectrum for the total gas density is due
to the gaseous spiral arms with a well-defined pitch angle $\varphi_\mathrm{m}$ of about $8^\circ$ with a HWHM of
$6^\circ$. We note that the pitch angles of the spiral arms in the total gas distribution show
somewhat less scatter (narrower spectral maximum) than those in CO. The difference between the
positions of the maxima in CO and the total gas is due to smaller pitch angles of spiral structures
in the outer galaxy, which are traced by HI, while CO traces spiral features with larger pitch
angles in the inner galaxy. A decrease in spiral pitch angles with increasing radius has been
observed in many galaxies \citep[e.g.][]{B07,beck15,VEBSF15}.

The lower panel of Fig.~\ref{fig-angles} shows clearly that the distribution of the
\textit{pitch angles of the structures} $\varphi_\mathrm{m}$
prominent in polarized intensity (magnetic arms) is
almost identical to the distribution of the CO structures: Most of the CO emission and polarized intensity are
localized in regions with $\varphi<30^\circ$.
The pitch angles of the polarization structures are only slightly larger (the peak pitch angles at $\varphi\approx14^\circ$) than those of the total gas.
It is notable that the
\textit{pitch angles of the magnetic field vectors} $\varphi_\mathrm{p}$
are concentrated at significantly different values,
$20^\circ$--$35^\circ$ with a
maximum at $\varphi\approx26^\circ$ and the HWHM of $15^\circ$.
This difference cannot be explained by Faraday rotation in the foreground
of the Milky Way, which is only about $-2^\circ$ at $\lambda6\cm$, as estimated from the
$B$ vectors at $\lambda6\cm$ and $\lambda13\cm$ (see Sect.~\ref{ATCA}).

To summarize, the peak pitch angles $\varphi_\mathrm{m}$ of the total gas
and magnetic spiral arms are very close to each other at about $10^\circ$,
whereas the magnetic field vectors have a different peak pitch angle $\varphi_\mathrm{p}$
of about $26^\circ\pm15^\circ$, where the ranges represent the HWHM of the spectral maxima.
The local differences between the pitch angles of the magnetic field vectors and the spiral arms can
still be quite large (as is obvious from Fig.~\ref{fig-magarms}).

\begin{figure}
\centering
\includegraphics[width=0.35\textwidth]{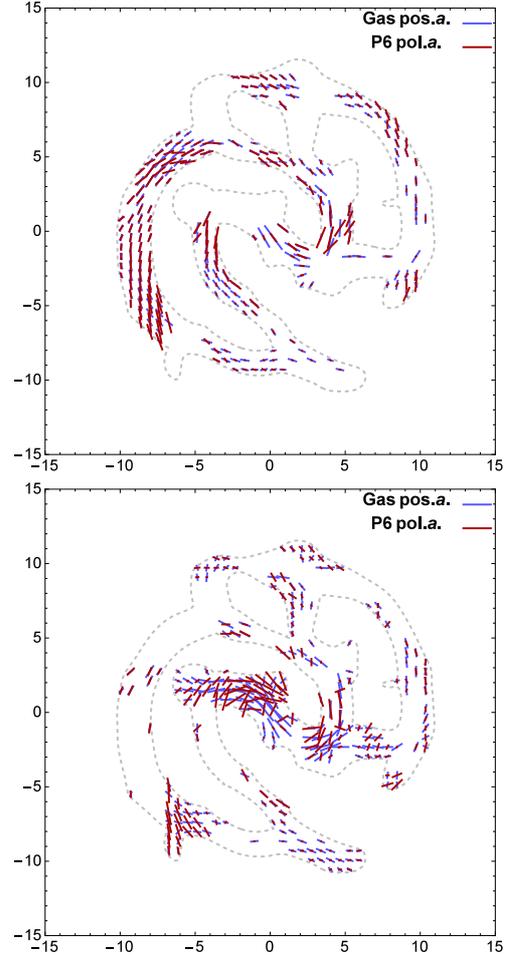}
\caption{\label{fig-matarms}Position angles of the elongated structures of the total neutral gas
(blue bars), obtained from the anisotropic wavelet transform on the scale $2a=1\arcmin$, and
magnetic polarization angles (red bars); the isocontour of the total gas intensity at 20\% of its
maximum in the arms is shown as a dashed line. The upper panel shows the data points where the
difference between two angles is between $3^\circ$ and $33^\circ$ with a positive
value meaning a clockwise rotation with respect to the gas structure), as suggested by the relative shift of the
corresponding peaks in the angular spectra (Fig.~\ref{fig-angles}), while the lower panel shows the
remaining data points. The axis scales are in kpc.}
\end{figure}
\begin{figure}
\centering
\includegraphics[width=0.35\textwidth]{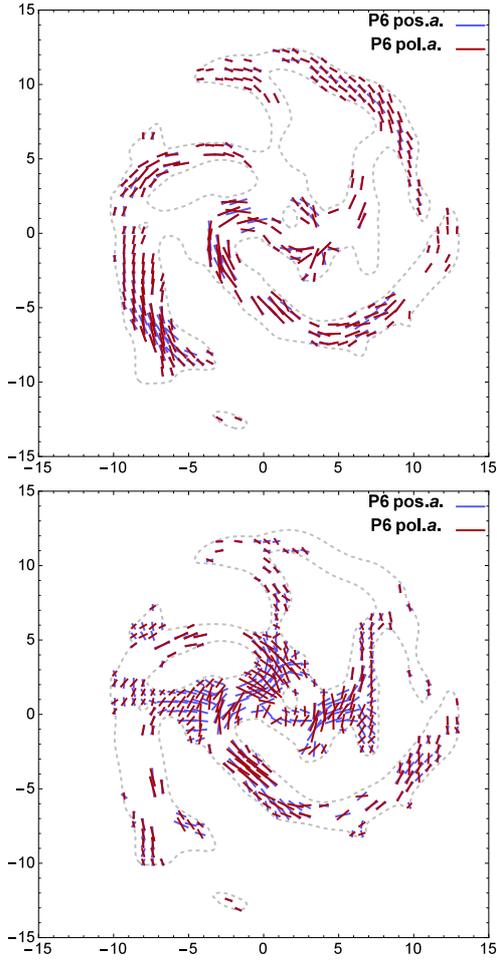}
\caption{\label{fig-magarms}As in Fig.~\ref{fig-matarms}, but with blue bars showing the position
angles of the structures of the polarized intensity P6, the isocontour of P6 at 20\% of its maximum
intensity shown as dashed.
The upper and lower panels show data points with the difference between the angles
in the range $-1^\circ$ to $29^\circ$ and out of this range, respectively.}
\end{figure}

To confirm the identification of the regions that contribute to the maxima in the angular spectra
of Fig.~\ref{fig-angles}, we show in Fig.~\ref{fig-matarms} the polarization vectors of the ordered
magnetic field position angles of the gas structures in the regions that contribute to the maxima
in their angular spectra. The upper panel shows the arm fragments where the
orientation
of the ordered magnetic field differs by $18^\circ\pm15^\circ$
from the local axis of a material arm, which is the case in most parts of Arms I and II.

The lower panel demonstrates that the two angles differ significantly and systematically in the
bar, apart from localized regions elsewhere in the galaxy. Such deviations occur where the position
angle of the gas arm turns sharply (arm sections Ib and IIb). This is a hint that different
mechanisms can be responsible for the magnetic field alignment along the material arms in different
parts of the galaxy.

\begin{figure}
\centering
\includegraphics[width=0.35\textwidth]{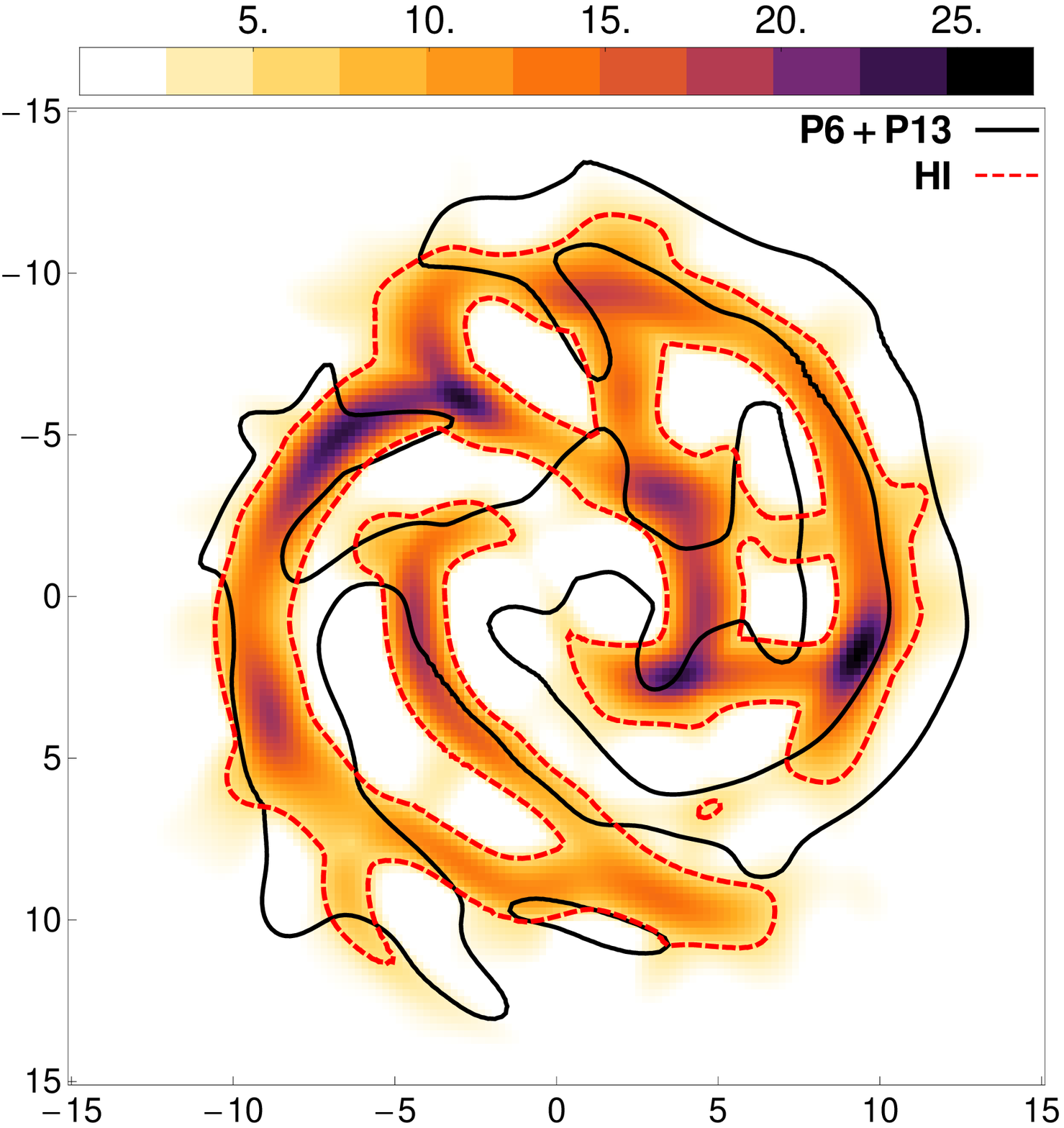}
\includegraphics[width=0.35\textwidth]{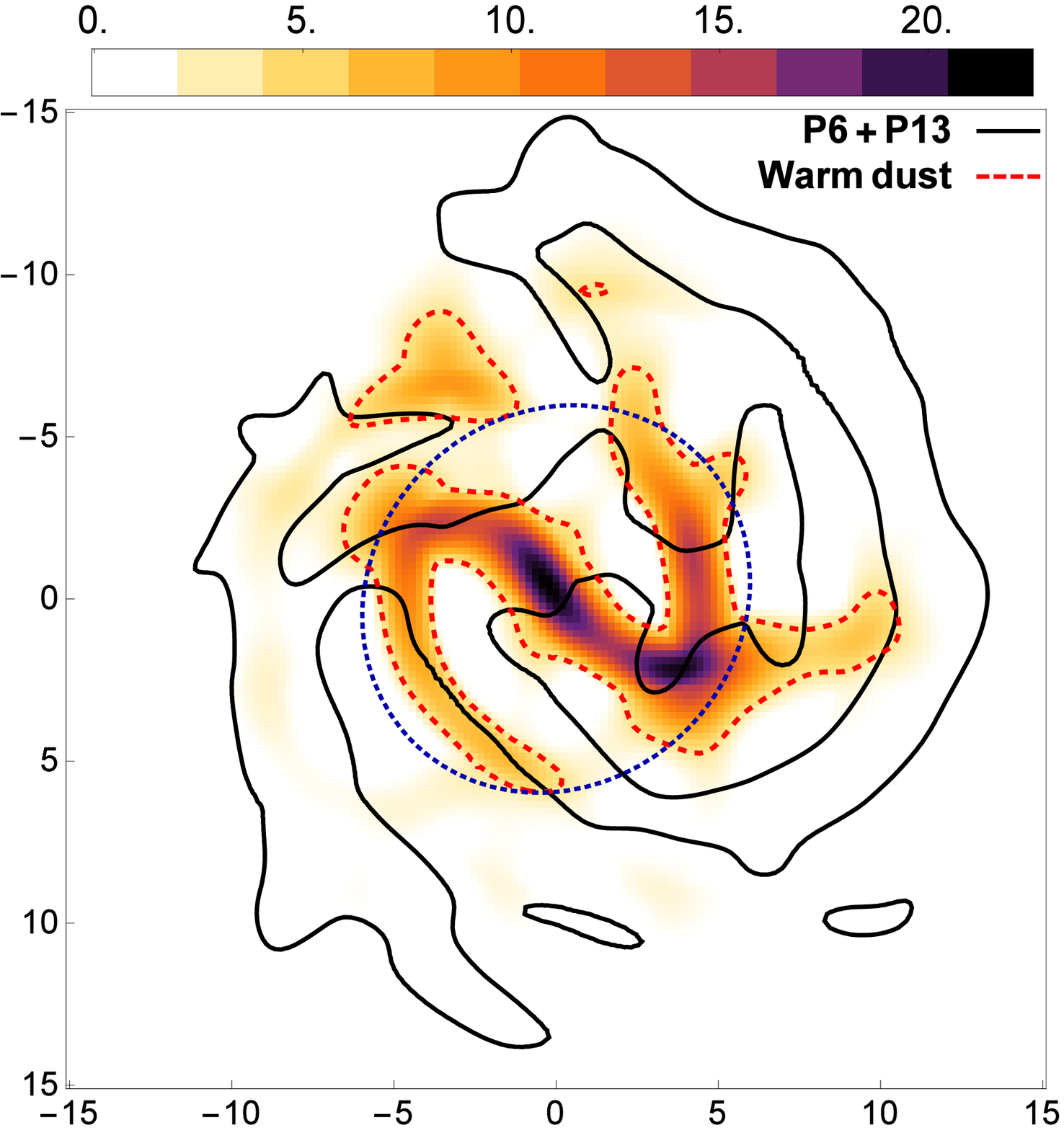}
\caption{\label{fig-fin}Locations of the elongated structures, on the scale
$2a=1\arcmin$, of the combined polarized radio emission (P6+P13) (black isocontour at 20\%
maximum intensity) and in the distributions of $\HI$ (top panel) and warm dust (bottom panel), respectively,
shown in colour and as dotted contours at 20\% of its maximum intensity. The axis scale is given in
kiloparsecs assuming the distance to M83 of $8.9\Mpc$. The dashed ellipse shows the corotation radius.}
\end{figure}

The analysis of the relative orientations of magnetic pitch angles and those of spiral arm segments is
complicated by the fact that the magnetic and material arms often do not overlap, even if parallel
to each other. In order to improve the comparison we take advantage of the fact, quantified above,
that the pitch angles of the spiral arms in CO and P6 are very similar to each other, and present
in Fig.~\ref{fig-magarms} the polarization and position angles obtained from the magnetic tracer
P6, which is the polarized intensity at $\lambda6\cm$. Now the area where the comparison is possible is
substantially larger than in the previous figure,
confirming that the ordered magnetic field is well aligned with the spiral arms along most
sections of the Arms I and II (as discussed above, the two orientations differ systematically by an
angle of about $10^\circ$ which is difficult to discern
visually), while there is little or no alignment in
the bar region.

The reason for the
misalignment in the bar region in Fig.~\ref{fig-magarms} (bottom) is a structure in P6 that appears on large
scales due to merging of smaller structures (Fig.~\ref{fig:cmap4p} bottom). The existence of a
smaller structure becomes clear from comparing P6 from Fig.~\ref{fig:maps} with P6 from
Fig.~\ref{fig:cmap4p}. In the
unfiltered map (Fig.~\ref{fig:maps}), P6 is concentrated in two narrow features with a position angle of about
$50^\circ$, running roughly parallel to the bar
visible in I6, gas, and dust, with the polarization angles
(Fig.~\ref{fig:pi6})
roughly aligned with them. At both ends of the bar, the position angles of the structures, as
well as the polarization angles, jump by about $70^\circ$, to become oriented roughly in the north-south
orientation. Most of these features are lost in Fig.~\ref{fig:cmap4p} because the wavelet transform
averages over small structures. In Fig.~\ref{fig:cmap4p} bottom, the strongest wavelet structure
near the centre is
a combination of many narrow features in the original image and no longer resembles the bar.
Not surprisingly, the structure position angles are between $80^\circ$ and $130^\circ$, so completely
different from those of the polarization angles.

$\HI$ and the combination of polarized emission at $\lambda6\cm$ and $\lambda13\cm$ cover the
largest part of the galaxy and thus reveal the most extended spatial distributions. Only these two
maps allow us to identify both arms along their whole lengths. Other tracers contain lacunae and
often exhibit only short segments of arms, especially in the inner galaxy (e.g. in the $\HII$ map).
The wavelet-filtered image of $\HI$ and an isocontour of combined polarized radio emission, P6+P13,
are shown in Fig.~\ref{fig-fin} on the scale $2a=1\arcmin$. The spiral arms are very visible in
both tracers.
In $\HI$, there are two bridges with large pitch angles (but not radial), which are seen in both tracers
(Fig.~\ref{fig:maps}). The figure supports the conclusion that all tracers show Arm II at the same
positions, while the positions of Arm I display a systematic relative shift: the magnetic arm is
shifted with respect to the $\HI$ arm, inwards in the inner galaxy (arm segment Ia) and outwards
in the outer galaxy (Ib and Ic).

\section{Discussion}\label{discussion}

A traditional approach to quantifying galactic spiral structures and estimating their pitch angles is
based on the Fourier transform of the light (or any other suitable variable) distribution in
azimuth \citep[][and references therein]{PD92}. Such a spectral analysis involves a fit to an
average pitch angle for a whole galaxy assuming the logarithmic-spiral shape of the arms. Given the
complex structure of the spiral patterns with numerous branches and strong spatial variations of the
pitch angle, etc., even in galaxies with grand-design patterns such as M51, the global approach to
quantifying the spiral structure is rather restrictive. \citet{PEB14} attempted to improve
the method in this respect by applying it to relatively narrow annuli rather than to a galaxy as a
whole. However, a better way to resolve these problems would be to employ wavelet transforms, as we
do here \citep[see also][]{frick00,frick01,patrikeyev06}. The wavelet-based approach is free of any
model assumptions about the shape of the spiral arms, and their segments (such as a perfect
$m$-armed pattern or a logarithmic spiral) allows for different widths of arms and inter-arm
regions and works equally well for any structures, whether spiral, azimuthal, or radial.

\subsection{Magnetic arms}\label{MAD}

M83 has well-defined magnetic arms where the ordered magnetic field is concentrated, which is visible as
structures in polarized radio emission of about $1\arcmin$ in width. Similar to the galaxy
NGC\,6946, the polarized arms appear to be more or less independent of the spiral patterns seen in
other tracers, since they are displaced from the latter in large areas by about $20^\circ$ along the
azimuth on average in the mean shortest distance. This does not exclude a physical relation
between them, which is suggested by their close spatial proximity and the fact that their axes are
well aligned with each other. The mutual location of the gaseous and magnetic arms is diverse, with
magnetic Arms Ib, Ic, and IIa extended along the inner edge of the corresponding gaseous arms, but
magnetic Arm IIb and the inner part of Arm IIc superimposed on the corresponding gaseous arms
(Figs.~\ref{fig:cmap4g}, \ref{fig:cmap4p}, and \ref{fig-fin}).

This diversity suggests that a variety of physical mechanisms are responsible for the formation of
the magnetic arms, possibly acting simultaneously. A remarkable feature of the large-scale galactic
magnetic field that manifests itself as magnetic arms displaced from the gaseous arms is that its
strength is greater where the gas density is lower. This may imply that the large-scale galactic
magnetic fields are not frozen into the interstellar gas and thus need to be continuously
replenished, presumably by the galactic mean-field dynamo action \citep{S07}. The only apparent
alternative is that magnetic arms represent slow MHD density waves \citep[][and references
therein]{Lou1998,LF03,LB06}. However, mean-field dynamo equations in a thin disk, either linear or
nonlinear, do not admit wave-like solutions under realistic conditions even when generalized to
include the second time derivative of the magnetic field (the telegraph equation)
\citep{Chamandy2013,Chamandy2014}. It cannot be excluded that more advanced nonlinear dynamo
models, especially those with the direct coupling of the induction and Navier--Stokes equations,
will reveal a new class of dynamo solutions reminiscent of the slow MHD waves.

There are several physical processes that can contribute to the formation of magnetic arms
displaced from the gaseous arms:
\begin{enumerate}
\item The mean-field dynamo
action can be suppressed within the gaseous arms by either a presumably enhanced
fluctuation dynamo driven by stronger star formation \citep{Moss2013} or a stronger galactic
outflow driven by stronger star formation \citep{SSS07,Chamandy2015} \citep[see
also][]{Shukurov1998}.

A generic problem of such mechanisms is that they produce the desired displacement only within a
few kiloparsecs of the corotation radius of the spiral pattern
\citep{Shukurov1998,Chamandy2013,Chamandy2014}, because the residence time of a volume element
within a gaseous spiral arm is shorter than the dynamo time scale of order $5\times10^8\yr$ at
large distances from the corotation.
 {A few} mechanisms, briefly discussed below, have
been suggested
to produce magnetic arms displaced from the gaseous ones far away from the corotation radius.

\item  {The model by \citet{Moss2013,Moss2015}
assumes that a large-scale regular field is generated everywhere
in the disk, while a small-scale dynamo injects turbulent fields only in the spiral arms. This gives
polarization arms between the gaseous arms at all radii, but with pitch angles of the polarization structures
and pitch angles of the polarization vectors that are significantly smaller than those of the gaseous arms, in contrast
to the observations discussed in this paper.}

\item A finite mean-field dynamo relaxation time (a temporal non-locality of the mean electromotive
force) does produce, under realistic parameter values, a displacement of up to $30^\circ$ in
azimuth at the corotation radius, with magnetic arms lagging behind the gaseous ones
\citep{Shukurov1998,Chamandy2013}. The ridges of magnetic arms thus produced have a systematically
smaller pitch angle than do the gaseous spirals because of the action of differential rotation
on the large-scale magnetic field.
The pitch angle of the ordered magnetic field in M83 is, on average, larger than that
of the material arms, contrary to this model (which, however, has not been
specifically developed for M83).


\item A different approach to the problem of displaced magnetic and gaseous spiral patterns emerges
if the material pattern is not a solidly rotating structure, as suggested by the density wave
theory, but rather a transient and evolving system of spiral arm segments produced by a bar,
galactic encounters, local instabilities, etc., and wound up by the galactic differential rotation.
As shown by \citet{Chamandy2013,Chamandy2014,Chamandy2015}, the mechanisms mentioned above are
relieved of their problems in this case, producing diverse interlaced or intersecting magnetic and
gaseous spiral patterns depending on the relative contribution of each mechanism.

\item Compression of the both the large-scale and turbulent magnetic fields in the gaseous arms
enhances polarized synchrotron emission within them, also by amplifying the anisotropy of the
random magnetic field \citep[see Section~8 of][]{Becketal05,patrikeyev06,fletcher11}. The resulting
ridges of enhanced polarized emission are expected to be located at the inner edge of the material
arms inside the corotation radius and at the outer edge outside the corotation. Offsets between the
spiral arm ridges in various tracers, of a few 100\,pc, are predicted by the density wave theory
\citep{R69} and were detected in a wavelet analysis of M51 data \citep{patrikeyev06}. Compression
of a turbulent field aligns the magnetic arm along the shock front, while compression of an ordered
field changes the pitch angle of the polarization angle to become more similar (but not identical)
to the pitch angle of the material arm.
\end{enumerate}

Most of these mechanisms (except those of \citealt{Moss2013,Moss2015})
are sensitive to the location of the corotation radius since advection of
magnetic fields from the gaseous arms would unavoidably affect the relative positions of the
magnetic and gaseous arms.
It has been suggested  that the corotation radius in M83 is about
$2.3\arcmin$--$2.4\arcmin$ \citep{KL91,RLH99,HKBEHMTNK14}, about $6\kpc$ at a distance of 8.9\,Mpc.
As usual in barred galaxies, this radius is only slightly larger than the major axis of the bar.
This does not allow us to clarify the role of the corotation radius in the formation of magnetic
arms using M83 as an example.

The magnetic Arm IIb--IIc coincides with the corresponding material arm (Fig.~\ref{fig-fin}). This
may still be consistent with the mechanisms (3) or (4).
The similarity of the magnetic pitch angles as
compared to the pitch angles of the gas arms (Figs.~\ref{fig-angles} and \ref{fig-matarms}) do not
lend support to the mechanism (2).
A regular field with a spiral pattern generated by a mean-field dynamo is possibly compressed
and partly aligned in density waves interacting with the magnetic field. Another possibility may be
associated with a coupling between magnetic fields and density waves.

Several narrow spiral arm segments are prominent in the dust emission in the inner galaxy
(Fig.~\ref{fig:maps}), which is indicative of compression. These features are
visible in the anisotropic wavelet transform maps on the scale $0.7\arcmin$ (Figs.~\ref{fig:cmap4g}
and \ref{fig:cmap4p}, top panels). The features in total neutral gas and ordered magnetic field
coincide with them in the west and north of the galaxy. Density wave models predict an offset of a
few 100\,pc between such features, corresponding to $0.1\arcmin$, which cannot be resolved with the
present observations. However, \citet{patrikeyev06} find evidence of such offsets in the galaxy
M51.

The magnetic Arm Ia--Ib is displaced from the corresponding material arm (Fig.~\ref{fig-fin}). Its
properties could be consistent with the mechanism (4) if the corotation radius of M83 is located at
about 7\,kpc at the assumed distance \citep{Lundgren04}, and this is apparently the case. However, the
displacement is too large to be consistent with density-wave compression. Alternatively, the whole of
Arm I may be generated by the mean-field dynamo via the mechanisms (1)--(3).
A rather good alignment of the polarization angles with the position angle is seen in the magnetic Arm I (Fig.~\ref{fig-magarms}, top panel). Such an alignment is facilitated by the mechanism (4).

An additional aspect of the magnetic field configuration in M83 is related to the presence of the
bar. Although the bar of M83 is shorter (about 7\,kpc) than that of NGC\,1097 (of about 16\,kpc),
the ordered magnetic fields are inclined with respect to the bar in both galaxies
(Fig.~\ref{fig-matarms}). The behaviour of polarization angles in the bar of M83 looks similar to
the behaviour in NGC\,1097 and NGC\,1365, which are other barred galaxies \citep{Becketal05}, and
indicates a non-axisymmetric gas flow in the bar region.
Indeed, Fig.~\ref{fig:pi6} shows that the magnetic field is inclined to the bar
axis in the regions SE and NW of the bar (at $x \approx -2', y\approx 0'$ and $x\approx +3', -2' < y < +2'$),
seen also in the bottom left of Fig.~\ref{fig:pmap1}, though less well owing to the wavelet smearing. Furthermore,
the ordered magnetic field near the bar is concentrated in two narrow regions where the magnetic field vectors are
aligned along the bar on the downstream (outer) sides of the bar (at $x\approx 0', y\approx +2'$ and $y\approx -1'$).
This is most visible in Fig.~\ref{fig:pi6}, but hardly in Fig.~\ref{fig:pmap1}. This whole pattern is similar
to what is observed in NGC\,1097 and NGC\,1365, though on a smaller scale.
The inner region of IC\,342 also shows magnetic fields strongly inclined to the bar \citep{beck15}.

\subsection{Magnetic field alignment with spiral arms}

The application of anisotropic wavelet transform provides an excellent opportunity to estimate the
pitch angles of various spiral structures without any model assumptions and to obtain their
quantitative characteristics in a controlled manner. In particular, the estimates of the pitch
angles from the angular spectra presented in Section~\ref{ORMFSA} are perhaps the most reliable
amongst the existing estimates of this kind. On average, the pitch angle of the magnetic field vectors
at $\lambda6.2\cm$ (presumably, mostly the regular field contaminated by an anisotropic random part)
is larger than that of the material spiral arms by about $20^\circ$: The magnetic field lines
form slightly more open spirals than the material arms. \citet{VEBSF15} present a compilation of the
measurements of magnetic and spiral-arm pitch angles in a sample of nearby galaxies (their Fig.~\ref{fig:pmap1}
and the accompanying text) and conclude that they are closely correlated and yet gives
a typical difference in their data of about $5^\circ-10^\circ$, with the
magnetic lines being more open than the spiral arms.
M83 generally fits this relation.

\section{Conclusions}

Magnetic structures and their inter-relations with the material structures are more complicated in
M83 than in NGC\,6946 and M51, where they were previously investigated in some detail.
The polarization angles in the vicinity of the {\em bar}\ of M83 are strongly inclined with
respect to the orientation of the bar, possibly indicating non-axisymmetric gas flows, as
also observed in other barred galaxies.
The relative intensities of {\em spiral structures}\ at various scales,
quantified by the wavelet power spectra, are similar
for the dust, gas, and total magnetic field. The pattern of the total radio continuum emission,
mainly associated with small-scale (turbulent) magnetic fields, looks more or less similar to the
material pattern. On the other hand, ordered magnetic fields (traced by polarized radio
emission) are distributed differently, showing weak correlation with other tracers.

We have identified two main material spiral arms that start at the ends of the bar. M83 also has
well-defined magnetic arms where the ordered magnetic field is concentrated. The magnetic arms are
partially displaced from the corresponding material arms (as in NGC\,6946), while they partially
coincide with the material arms in other regions {similar to M51).

The magnetic Arm IIb--IIc (see Fig.~\ref{fig-arms} for the notation) coincides with the material
arm, so behaves similarly to the outer south-western arms of M51 \citep{patrikeyev06}. The
polarization angles are mostly aligned with the orientations of the main material Arm II, but
generally have pitch angles larger by about $20^\circ$. This could be a signature of a spiral field
generated by the mean-field dynamo that is aligned by a still unknown mechanism, possibly
interacting with density waves. Major deviations occur in two regions where the gas arms have a
sharp bend (north-east and south-west of the bar), which the magnetic field does not follow.
Figure~\ref{fig:pi6} shows that the pattern of magnetic vectors at the bar ends in the north-east and
south-west is smoother than the structure of gas and dust, which is not an effect of limited
resolution.

The most prominent magnetic arm is displaced inwards with respect to the material Arm I in the
south (Ia), located in the inter-arm space between the corresponding material arm and the bar
(Fig.~\ref{fig-fin}), resembling the magnetic arms in NGC\,6946. The magnetic arm continues to the
west and north, displaced outwards with respect to the material Arm Ic. The approximate alignment
of the polarization angles with the magnetic arms (Fig.~\ref{fig-magarms}) was also found for the
magnetic arms in NGC\,6946.

The two types of magnetic arms may be different in their generation.
A displacement can indicate generation by the mean-field dynamo, while coincidence could be
due to shear or compression by density waves. A regular field generated by a mean-field
dynamo can be compressed in material arms and thus be partly aligned with them.
Interaction with nearby dwarf galaxies in the M83 galaxy group \citep{karachentsev07} may also
affect the gas kinematics and the magnetic field pattern; the dwarf galaxy NGC~5253 is possibly causing
the warp in the $\HI$ disk \citep{rogstad74}. The coexistence of different generation mechanisms could be the
result of the different timescales involved: While the dynamo operates on timescales of several rotation
periods, density waves or disturbances by interaction act on shorter timescales.
Dynamo models have indicated that gas streaming due to an encounter can enhance total and polarized emission
without any increase in the regular magnetic field strength \citep{Moss14}.

In summary, the modulation of the galactic dynamo by a transient spiral pattern is a
promising mechanism for producing such complicated spiral patterns as in M83.

A better understanding of the structure and origin of the magnetic arms observed in M83 needs a
high-resolution map of Faraday rotation to allow one to distinguish regular magnetic fields from the
anisotropic magnetic fields produced by compression and shear in the bar and spiral arm regions.
The radio polarization data presented in this paper are insufficient to provide reliable rotation
measures because Faraday depolarization is strong in the inner galaxy at $\lambda13\cm$.
Spectro-polarimetric observations and application of RM Synthesis \citep{brentjens05} are desirable, e.g.
with VLA data in the L band (1--2\,GHz) and S band (2--4\,GHz) giving a resolution in
Faraday depth of $\simeq 40\radm$. M83 is also proposed as a prime target
for the Square Kilometre Array (SKA) \citep{Becketal15}.

\begin{acknowledgements}
We thank Stuart Ryder for providing the H$\alpha$ images of M83.
We thank Sui Ann Mao and the anonymous referee for useful suggestions that helped to
improve the paper.
RB acknowledges support from the DFG Research Unit FOR1254. AS is grateful to STFC (grant ST/L005549/1)
and the Leverhulme Trust (grant RPG-2014-427) for their support.
PF, RS, and DS thank Prof.\ Michael Kramer for supporting several visits to the MPIfR.
\end{acknowledgements}

\bibliographystyle{aa}
\bibliography{rm}

\end{document}